\documentclass[12pt]{amsproc}

\usepackage[margin=1in]{geometry}
\usepackage{amsfonts}
\usepackage{amssymb}
\usepackage{mathrsfs}
\usepackage[dvips]{graphics}
\usepackage{epsfig}
\usepackage{euscript}
\usepackage{color}
\usepackage{bbm}
\usepackage[colorlinks ,pdfstartview=FitB]{hyperref}
\hypersetup{allcolors=blue}

\usepackage{hyperref}

\newtheorem{thm}{Theorem}[section]

\theoremstyle{definition}

\theoremstyle{remark}

\numberwithin{thm}{section}
\numberwithin{equation}{section}

\newcommand{\R}{{\mathord{\mathbb R}}}
\newcommand{\N}{{\mathord{\mathbb N}}}

\newcommand{\C}{{\mathord{\mathbb C}}}
\newcommand{\Z}{{\mathord{\mathbb Z}}}
\newcommand{\E}{{\mathord{\mathbb E}}}

\def\idty{{\mathchoice {\mathrm{1\mskip-4mu l}} {\mathrm{1\mskip-4mu l}} %
{\mathrm{1\mskip-4.5mu l}} {\mathrm{1\mskip-5mu l}}}}

\DeclareMathOperator{\tr}{tr}
\DeclareMathOperator{\Tr}{Tr}

\DeclareMathOperator{\supp}{supp}
\DeclareMathOperator{\diag}{diag}

\usepackage{hyperref}
\hypersetup{
    colorlinks=true,
    linkcolor=cyan,
    filecolor=magenta,      
    urlcolor=blue,
}

\begin{document}

\title[Disordered Quantum Spin Chains]{Aspects of the Mathematical Theory of\\ Disordered Quantum Spin Chains}


\author{G\"unter Stolz}
\address{Department of Mathematics, University of Alabama at Birmingham, Birmingham, AL~35294-1170, USA}
\email{stolz@uab.edu}
\thanks{}


\date{}

\begin{abstract}
We give an introduction into some aspects of the emerging mathematical theory of many-body localization (MBL) for disordered quantum spin chains. In particular, we discuss manifestations of MBL such as zero-velocity Lieb-Robinson bounds, quasi-locality of the time evolution of local observables, as well as exponential clustering and low entanglement of eigenstates. Explicit models where such properties have recently been verified are the XY and XXZ spin chain, in each case with disorder introduced in the form of a random exterior field. We introduce these models, state many of the available results and try to provide some general context. We discuss methods and ideas which enter the proofs and, in a few illustrative examples, include more detailed arguments. Finally, we also mention some directions for future mathematical work on MBL.

\end{abstract}

\maketitle

\tableofcontents

\section{Introduction}

One-particle localization (or Anderson localization) and many-body localization (MBL) are fundamentally different concepts. Both can be understood as the absence of quantum transport. However, in the one-particle case transport necessarily means that the particle moves, while for a many-body system transport can happen in the form of a group wave, where all individual particles move very little (think of Newton's cradle, the toy from your high school physics lab). 

It is very natural to view the latter as the transport of information through a particle system.  And it is here where quantum spin systems (or systems of qubits) arise as an important class of examples, as MBL for these systems can be interpreted as a mechanism which prevents the flow of information through such systems. This is generally considered as undesirable and frequently associated with the presence of disorder (random impurities, dirt) in the system. But, on the positive side, it could also be helpful for keeping information localized in a quantum storage device.

There is also an important conceptional advantage in looking at quantum spin systems when trying to understand MBL, rather than at other particle systems. Spins are the most simple non-trivial quantum particle, described by a two-dimensional Hilbert space and without any spatial degree of freedom (as opposed to, say, electrons). Thus transport in a quantum spin system can only appear in the form of information transport between particles, exactly the phenomenon whose presence or absence we would like to understand. 

Our goal here is to give a (hopefully) pedagogical introduction into some aspects of the mathematical theory of disordered quantum spin chains and, in particular, their MBL properties, mostly by discussing the concrete examples of the disordered XY and XXZ chains. The restriction to chains, i.e., one-dimensional systems, is essentially forced due to the complete lack of results on MBL for multi-dimensional quantum spin systems. For the same reason we will have nothing to say about many-body delocalization, often described in the context of thermalization or the ergodic phase of a many-body system in the physics literature. To appreciate the mathematical difficulty behind describing and rigorously establishing a delocalized regime in disordered many-body systems, it should be kept in mind that there is not even a rigorous proof of the existence of extended states in the three-dimensional (one particle) Anderson model at low disorder. We will take this as an excuse for not touching on this topic in the many-body context.

As for more bad news, we will not attempt to give a sufficiently complete list of MBL related references in the physics and quantum information literature. We will mention a few of these works where it fits into the context of our presentation and otherwise refer to the recent review articles \cite{AbaninPapic} and \cite{Agarwaletal}, which have extensive lists of references, hoping that this can provide a starting point into further reading on this rapidly growing research field. 

After those disclaimers, let us describe the more humble goals which we will try to accomplish in the remaining sections of this work:

To motivate the objects which we will refer to as manifestations of MBL, we will start in Section~\ref{sec:fund} by recalling some known properties of (deterministic) quantum spin chains and, in particular, introduce the concepts of Lieb-Robinson bounds, exponential clustering of correlations and area laws for the bipartite entanglement of states. Section~\ref{MBLman} provides a first discussion on how the MBL phase of a disordered spin chain will be reflected through these concepts.

Section~\ref{sec:XY} gives a survey of results for the prototypical and most simple model of a quantum spin system where all these MBL manifestations can be proven rigorously, the disordered XY chain. That this is possible is due to the fact that the XY chain can be mapped to a free Fermion system via the Jordan-Wigner transform, as we will recall in Section~\ref{sec:XYproofs} together with some properties of free Fermion systems, such as quasi-free states and their characterization through two-point correlation matrices, which are relevant to the proofs.

Much of the rest of this account will be used to discuss some recent results on MBL in the disordered XXZ spin chain. More precisely, these results refer to the droplet regime of the XXZ chain in its Ising phase, an anisotropic version of the Heisenberg model where the $Z$-term dominates the interaction, as measured by choosing the relevant anisotropy parameter $\Delta>1$. We start this in Section~\ref{sec:Ising} with a look at the limiting Ising model where $\Delta$ becomes infinite. This model is essentially trivial by being exactly diagonalizable, but still gives a good first indication of the properties of the low energy droplet regime in the XXZ chain for $1<\Delta<\infty$. The latter will be introduced in Section~\ref{sec:XXZdet}, discussing, in particular, the consequences of particle number conservation in the XXZ model, followed by a presentation of how this regime gets many-body localized after exposure to a random field, see Section~\ref{MBLXXZ}.

These results on MBL for the droplet regime of the disordered XXZ chain have been proven in several recent works, see \cite{BeaudWarzel2017a, BeaudWarzel2017b, EKS2017a, EKS2017b}, covering a broad range of manifestations such as exponential clustering and an area law for eigenstates, a zero-velocity Lieb-Robinson bound and quasi-locality of the Heisenberg evolution of local observables. We will explain some of the main ideas and techniques behind these proofs and, in Section~\ref{sec:proofs}, include a few detailed proofs which we consider to be particularly illustrative. This is far from giving a full picture of the technical intricacies involved in the localization proofs, but it will provide at least one example of how Schr\"odinger operator techniques enter (to show the Combes-Thomas bound in Section~\ref{sec:CT}) as well as another example illustrating the many-body techniques involved (for the special case of exponential clustering shown in Section~\ref{proofexpcl}). Our main reason for this choice is to highlight that ideas from these two worlds complement one another in the proofs.

In the concluding Section~\ref{sec:problems} we discuss some possible directions for future work, without claiming to provide a to-do-list which is anywhere close to complete. This will range from open questions for the concrete models we have focused on here, to other models whose study we consider promising, and to some of the broader goals which the mathematical theory of MBL will eventually have to address. This leaves many major challenges and will require a lot of new ideas. But we are optimistic that substantial further progress in the field of disordered quantum many-body systems can be achieved.

\vspace{.3cm}

{\bf Acknowledgements:} This article is an expanded version of the contents of several  mini-courses given by the author in 2017 and 2018. Presentation slides for all of these lectures are available electronically:

\begin{itemize} 
\item For the Master Class on Exotic Phases of Matter at the University of Copenhagen, 15 to 19 May 2017, at
\begin{center} \href{url}{https://qmath.ku.dk/events/conferences/masterclass2017/titles-and-abstracts/} \end{center}
\item For the Workshop and Summer School on Mathematical Aspects of Disordered Systems at ETH Z\"urich, 29 May to 2 June 2017, at
\begin{center} \href{url}{http://www.itp.phys.ethz.ch/research/mathphys/graf/mads/program.html} \end{center}
\item For the Arizona School of Analysis and Mathematical Physics at the University of Arizona, 5 to 9 March 2018, at
\begin{center} \href{url}{http://math.arizona.edu/$\sim$mathphys/AZSchool18/notes.html} \end{center}
\end{itemize}
I am grateful to the organizers of these Schools for providing such opportunities to young researchers and for giving me the chance to lecture. Without the latter these notes would never have been written. I also thank my co-authors Houssam Abdul-Rahman, Jacob Chapman, Alexander Elgart, Christoph Fisch\-bacher, Eman Hamza, Abel Klein, Bruno Nachtergaele and Robert Sims for a decade full of learning and discussions.

\section{Some fundamental properties of deterministic quantum spin chains} \label{sec:fund}

We start by surveying some key properties of deterministic quantum spin chains, which will serve to motivate our later discussion of MBL properties in spin chains with disorder. Here we will not attempt to present the most general known results and will instead focus on special cases which are sufficient for this purpose. We consider finite chains $[1,L] := \{1,2,\ldots,L\}$ of arbitrary length $L\in\N$ with a $1/2$-spin at each site, meaning that the Hilbert space is given by $\mathcal{H}_{[1,L]} = \bigotimes_{j=1}^L \C^2$. We will not discuss properties of the thermodynamic limit $L\to\infty$ and instead focus on bounds which are uniform in the finite volume $L$.

For simplicity, let us look at a Hamiltonian which consists of next-neighbor interactions $h_{j,j+1}$ between the sites $j$ and $j+1$, for $j=1,\ldots,L-1$, and local field terms $t_j$ acting at site $j$, $j=1,\ldots,L$, 
\begin{equation} \label{Ham}
H_L = \sum_{j=1}^{L-1} h_{j,j+1} + \sum_{j=1}^L t_j.
\end{equation}
For convenience we also assume that there is a constant $C<\infty$ such that $\|h_{j,j+1}\| \le C$, $\|t_j\|<C$ uniformly in $j$ (for the field terms this is not necessary in the results we state below). We do {\it not} assume translation invariance, i.e., that all operators $h_{j,j+1}$ are induced by the same operator $h$ on $\C^2 \otimes \C^2$ (and similar for $t_j$).

The velocity of group transport (or information transport) within a spin chain can be expressed in terms of Lieb-Robinson bounds. Once again for simplicity, we will formulate this for one-site obervables only, i.e., observables 
\begin{equation} 
A_j = I \otimes \ldots \otimes I \otimes A \otimes I \otimes \ldots \otimes I
\end{equation} 
acting non-trivially only on the $j$-th component of the tensor product $\mathcal{H}_L$ as a matrix $A\in \C^{2\times 2}$. By $\tau_t(A_j) = e^{it H_L} A_j e^{-itH_L}$ we denote the Heisenberg dynamics of $A_j$ under $H_L$.

\begin{thm} \label{LR}
There exist finite positive constants $C$, $\mu$ and $v$ such that
\begin{equation} \label{classicalLR}
\| [ \tau_t(A_j), B_k] \| \le C \|A\| \|B\| e^{-\mu(|j-k|-vt)}.
\end{equation}
for all $A, B \in \C^{2\times 2}$, $L\in \N$ and $1 \le j,k \le L$.
\end{thm}

The constant $v$ is interpreted as an upper bound on the group velocity within the spin system. The first result of this type was provided by Lieb and Robinson in \cite{LiebRobinson1972}, which resulted in the description of estimates of the form \ref{classicalLR} as {\it Lieb-Robinson bounds}. More recent important break throughs on Lieb-Robinson bounds and their applications were made in \cite{NachtergaeleSims2006} and \cite{HastingsKoma2006}. A survey with some of the most general result is in \cite{NachtergaeleSims}. These results are not restricted to one-dimensional systems.

Two other results which we want to recall here refer to the ground state of {\it gapped} spin systems, where we continue to assume the basic form \eqref{Ham}. In the simplest setting, being gapped means that the Hamiltonians $H_L$, $L\in \N$, have a non-degenerate normalized ground state $\varphi_L$ at energy $E_L$ and that 
\begin{equation} \label{gapped}
\min (\sigma(H_L) \setminus \{E_L\}) - E_L \ge \gamma > 0,
\end{equation}
uniformly in $L\in \N$.

\begin{thm} \label{ExpCl}
The ground state $\varphi_L$ of a gapped quantum spin system $H_L$ has exponentially decaying correlations, in the sense that there exist constants $C$ and $\mu>0$ such that
\begin{equation}
\mbox{Cor}_{\varphi_L}(A_j, B_k) := | \langle \varphi_L, A_j B_k \varphi_L \rangle - \langle \varphi_L, A_j \varphi_L \rangle \langle \varphi_L, B_k \varphi_L \rangle| \le C\|A\| \|B\| e^{-\mu|j-k|}
\end{equation}
for all $A, B \in \C^{2\times 2}$, $L\in \N$ and $1\le j,k \le N$.
\end{thm}

This {\it exponential clustering} property of gapped spin systems has been derived as an application of Lieb-Robinson bounds of the form \eqref{classicalLR}, e.g.\ \cite{NachtergaeleSims2006, HastingsKoma2006}. Another general property of gapped spin systems is an area law for the entanglement entropy of the ground state. To state this, let $1\le \ell < L$ and consider the bipartite decomposition 
\begin{equation} \label{bipartite}
\mathcal{H}_L = \mathcal{H}_A \otimes \mathcal{H}_B, \quad \mathcal{H}_A = \bigotimes_{j=1}^\ell \C^2, \quad \mathcal{H}_B = \bigotimes_{j=\ell+1}^L \C^2.
\end{equation} 
The bipartite entanglement entropy of a pure state $\rho_\varphi = |\varphi\rangle \langle \varphi|$ ($\phi \in \mathcal{H}_L$, $\|\varphi\|=1$) with respect to this decomposition is 
\begin{equation} \label{entangle}
\mathcal{E}(\rho_\varphi) = \mathcal{S}(\rho_\varphi^A) = -\Tr \rho_\varphi^A \log \rho_\varphi^A, 
\end{equation} 
with the reduced state $\rho_\varphi^A = \Tr_B \rho_\varphi$ and the von Neumann entropy $\mathcal{S}(\cdot)$.

\begin{thm} \label{AreaLaw}
The entanglement entropy of the ground state $\varphi_L$ of a gapped spin system satisfies an area law, i.e.,
\begin{equation}
\mathcal{E}(\rho_\varphi) \le C < \infty
\end{equation}
uniformly in $L\in \N$ and $1\le \ell < L$.
\end{thm}

This was first proven in \cite{Hastings}. It is an open problem whether an area law holds in $d$-dimensional gapped spin for $d\ge 1$, i.e., if there is a bound of the form $\mathcal{E}(\rho_\varphi) \le C\ell^{d-1}$ for the bipartite entanglement of a gapped ground state with respect to a box of sidelength $\ell$ within a box of sidelength $L$ in $\Z^d$.

\section{Understanding the MBL phase in light of Theorems~\ref{LR}, \ref{ExpCl} and \ref{AreaLaw}} \label{MBLman}

Localization in non-interacting quantum systems, e.g.\ Anderson localization, is generally understood either as localization (rapid decay) of eigenfunctions or as dynamical localization, e.g.\ the absence of quantum transport. A similar basic distinction can be made among the properties of interacting systems which are interpreted as manifestations of many-body localization.

\begin{itemize}

\item{\bf Dynamical MBL:} One interpretation of MBL is dynamical and is characterized by the absence of group transport (or information propagation). A strong form of this is given by a {\it zero-velocity Lieb-Robinson bound}, i.e., a bound of the form \eqref{classicalLR} with $v=0$. One can also consider the weaker condition where the exponential factor in \eqref{classicalLR} is replaced by $e^{-\mu(|j-k| - t^{\alpha})}$ for some $\alpha<1$. This would exclude ballistic many-body transport, but still allow for anomalous forms of transport, e.g.\ diffiusion for $\alpha=1/2$. We will concentrate on zero-velocity LR bounds here, but see \cite{Damaniketal} for examples of systems with anomalous many-body transport. Borderline situations for fields with decaying randomness were investigated in \cite{GebertLemm}.

\vspace{.3cm}

\item{\bf Eigenstate MBL:} The addition of disorder to a gapped spin system will typically close the ground state gap, or, more precisely, let its size tend to zero for large $L$ with high probability. Nevertheless, if the system is in an MBL phase, one expects that the ground state still exhibits exponential clustering of correlations and an area law for the bipartite entanglement entropy as in Theorems~\ref{ExpCl} and \ref{AreaLaw}, respectively. In fact, MBL properties such as these should extend to eigenstates at energies above the ground state energy, which is often expressed as the existence of a mobility gap in the system (but not a gap). In this language, a system can be called {\it fully many-body localized} if the mobility gap covers the entire spectrum. In other systems, one can try to show the existence of a many-body localization-delocalization transition by showing that there is a ``thermalized'' energy regime above the mobility gap (which is most pragmatically defined as an energy regime where none of the agreed upon characteristics of MBL hold).

\end{itemize}

Almost the entire remainder of this work will discuss two particular examples of disordered quantum spin chains in which these MBL properties have been verified rigorously. We start by surveying results for the XY chain in random transverse field (i.e., in the $Z$-direction). This provides a simple model where full many-body localization can be shown by directly reducing the MBL properties to localization properties of the non-interacting Anderson model. Then we include a more detailed discussion of MBL in the low energy droplet regime of the Ising phase XXZ chain in random $Z$-field. While we are not aware of any rigorous results on the existence of a thermalized phase in this model, it is physically expected that the disordered XXZ chain exhibits a localization-delocalization transition.

In addition to the three MBL manifestations introduced above, zero-velocity LR bounds, exponential clustering and area laws, we will mention results on related properties, including quasi-locality of the dynamics of local observables (Theorem~\ref{thm:XXZdyn} below) and dynamical area laws under quantum quenches (Theorem~\ref{thmdynent}). But our collection is certainly not exhaustive with respect to the range of physical phenomena which can be associated with MBL. In particular, we will not discuss the existence of a complete set of local integrals of motion (e.g.\ \cite{Serbynetal, Imbrieetal}) which is now physically accepted as the correct way to characterize the regime of full many-body localization.

\section{MBL Properties of the Disordered XY Chain} \label{sec:XY}

\subsection{The XY Chain and its Reduction to Free Fermions}

The isotropic XY chain (or XX chain) in random transversal field is given by the Hamiltonian
\begin{equation} \label{XYchain}
H_{XY}= H_{XY}(\omega) = - \sum_{j=1}^{L-1} (\sigma_j^X \sigma_{j+1}^X + \sigma_j^Y \sigma_{j+1}^Y) - \sum_{j=1}^L \omega_j \sigma_j^Z 
\end{equation}
in ${\mathcal H}_L = \bigotimes_{j=1}^L \C^2$. Here $\sigma_X$, $\sigma_Y$ and $\sigma_Z$ are the standard Pauli matrices. Throughout we will assume that 
\begin{eqnarray} \label{distribution}
\mbox{$(\omega_j)_{j=1}^{\infty}$ are i.i.d.~random variables, with distribution $d\mu(\omega_j) = \rho(\omega_j)\,d\omega_j$} \\ \mbox{and bounded, compactly supported density $\rho$. \notag}
\end{eqnarray}

Ever since the ground breaking work of Lieb, Schultz and Mattis \cite{LSM}, which showed how to exactly diagonalize \eqref{XYchain} for the case of constant field $\omega_j=c$ for all $j$, the XY chain has served as a first test case in many works on quantum spin chains. This is due to the fact, as observed by Lieb, Schultz and Mattis, that the XY chain can be mapped to a free Fermion system via the Jordan Wigner transform. The latter refers to the operators
\begin{equation} \label{JordanWigner}
c_1 := a_1, \quad c_j := \sigma_1^Z \ldots \sigma_{j-1}^Z a_j, \quad j =2,\ldots,L, \quad a := \begin{pmatrix} 0 & 1 \\ 0 & 0 \end{pmatrix},
\end{equation}
which satisfy the canonical anti-commutation relations (CAR):
\begin{equation}
\{c_j, c_k^*\} = \delta_{jk} I, \quad \{c_j,c_k\} = \{c_j^*, c_k^*\}=0,
\end{equation}
meaning that they provide a full set of Fermionic modes in the $2^L$-dimensional space ${\mathcal H}_L$. 

The Hamiltonian can be rewritten as
\begin{equation} \label{quad}
H_{XY} = 2 c^* M c + E_0,
\end{equation}
with $E_0 = -\sum_j \omega_j$, operator-valued column and row vectors $c=(c_1, \ldots, c_L)^t$ and $c^*=(c_1^*,\ldots,c_L^*)$ and
\begin{equation} 
M = \begin{pmatrix} \omega_1 & -1 & & \\ -1 & \ddots & \ddots & \\ & \ddots & \ddots & -1 \\ & & -1 & \omega_L \end{pmatrix}.
\end{equation}
In the language of second quantization this can be expressed as the unitary equivalence
\begin{equation}
H_{XY} \cong 2 d\Gamma_a(M) + E_0 
\end{equation}
on the antisymmetric Fock space ${\mathcal F}_a(\ell^2([1,L]))$ (and $d\Gamma_a(M)$ denoting the Fermionic part of the full non-interacting many-body Hamiltonian governed by the one-body Hamiltonian $M$). Thus $M$ takes the role of an {\it effective one-particle Hamiltonian} for $H_{XY}$, which, up to unitary equivalence, fully governs $H_{XY}$. More precisely, by diagonalizing $M = {\mathcal O}^T \Lambda {\mathcal O}$ via an orthogonal matrix $\mathcal O$ and diagonal $\Lambda = \diag(\lambda_j)$ one further reduces the XY chain from \eqref{quad} to the free Fermion system
\begin{equation} \label{freeFerm}
H_{XY} = 2\sum_{j=1}^L \lambda_j b_j^* b_j + E_0
\end{equation} 
for the Fermionic modes $b_j$ given by the components of the vector $b = {\mathcal O}c$. This free Fermion system has a unique (up to a phase) normalized vacuum vector $\Omega \in \cap_j \ker b_j$ and and orthonormal basis of eigenvectors and corresponding eigenvalues of $H_{XY}$ is given by
\begin{equation} \label{varphialpha}
\varphi_\alpha = \Big( \prod_{j:\alpha_j=1} b_j^* \Big)\Omega, \quad E_\alpha = 2\sum_{j:\alpha_j=1} \lambda_j + E_0, \quad \alpha \in \{0,1\}^L.
\end{equation}

Thus, what has been accomplished via the Jordan-Wigner transform is a reduction of the diagonalization of the many-body Hamiltonian $H_{XY}$ (on a $2^L$-dimensional space) to the diagonalization of the effective one-particle Hamiltonian $M$ (on an $L$-dimensional space). In the case of constant field $\omega_j= \mbox{const}$, the diagonalization of $M$ can be carried out explicitly, leading to the exact solution of the constant field XY chain by Lieb, Schultz and Mattis. For the random case considered here we find $M$ to be the one-dimensional Anderson model, restricted to the finite interval $[1,L]$. While this is not exactly solvable, its qualitative properties are very well understood in the form of strong forms of Anderson localization. More precisely, the following {\it localization of eigenvector correlators} holds under the assumption \eqref{distribution} on the distribution of the random parameters $\omega_j$ made above:
\begin{equation}  \label{ecorloc}
\E \left( \sup_{|g|\le 1} |(g(M))_{jk}| \right) = \E \left(\sum_\ell |\varphi_\ell(j)||\varphi_\ell(k)| \right)  \le C e^{-\mu|j-k|} ,
\end{equation}
uniformly in $L$. Here $g(M) = \sum_\ell g(\lambda_\ell) |\varphi_\ell \rangle \langle \varphi_\ell|$ in terms of the eigenvalues $\lambda_\ell$ and eigenvectors $\varphi_\ell$ of $M$. From this one easily sees the equality of the first two expressions in \eqref{ecorloc}, where the second one takes the form of a correlator of eigenvectors. As a special case of the first expression one sees that \eqref{ecorloc} incorporates {\it dynamical localization} of the form
\begin{equation} \label{dynloc}
\E \left( \sup_{t\in \R} |(e^{-itM})_{jk}| \right) \le C e^{-\mu|j-k|}.
\end{equation}

The remaining task in establishing localization properties of the disordered XY chain lies in understanding if and how Anderson localization \eqref{ecorloc}, \eqref{dynloc} of the effective Hamiltonian leads to MBL of the many-body Hamiltonian. We will now survey several results which accomplish this task. Combined, these results can be read as saying that the disordered XY chain \eqref{XYchain} is a fully many-body localized quantum spin system.

While this clearly shows the relative simplicity of understanding MBL of the disordered XY chain (in reducing it to Anderson localization), we point to two issues which have to be dealt with in this reduction and in the proofs of the following results: (i) The Jordan-Wigner transform \eqref{JordanWigner} is non-local, so that tracking if locality properties are preserved under inverting this transformation needs some care. (ii) While the spin chain has commuting degrees of freedom, i.e., local observables at individual sites of the chain (such as the spin raising operators $a_j$), the Fermionic modes $c_j$ are anti-commuting, so that two different concepts of locality have to be reconciled. 

\subsection{A Survey of Results on the Disordered XY Chain} \label{XYsurvey}

Here we will cover results from \cite{HamzaSimsStolz2012, KleinPerez1990, SimsWarzel2016, PasturSlavin2014, AbdulRahmanStolz2015, ARNSS2016}. See also the survey paper \cite{ARNSS2017}.

We start with dynamical MBL in the form of a zero-velocity LR bound:

\begin{thm}[\cite{HamzaSimsStolz2012}] \label{thmLR}
There exist $C<\infty$ and $\mu>0$ such that
\begin{equation} \label{XYLR}
\E \left( \sup_{t\in\R} \| [\tau_t(A_j), B_k] \| \right) \le C \|A\| \|B\| e^{-\mu|j-k|}
\end{equation}
for all $L$, $1\le j, k \le L$, $A, B \in \C^{2\times 2}$.
\end{thm}

We note that here and in all the following results for disordered spin chains, averaging $\E(\cdot)$ over the random parameters is used. Often, one could also formulate related bounds which hold with high probability (in particular, with probability tending to one in the infinite volume limit $L\to\infty$). While results on averages allow for the technically most elegant statement, we stress that this is not equivalent to almost sure deterministic bounds (the least to expect is that constants in such bounds will depend on the disorder). Also, from a probabilistic point of view bounds on averages are only a first step to understand random variables. Additional understanding of fluctuations is often desirable to provide further insight, but we will not make any contributions to this here (one example where fluctuations have been considered is \cite{EPS}, in this case for entanglement bounds in disordered free Fermion systems). 

We proceed with a result on exponential clustering.

\begin{thm}[\cite{SimsWarzel2016}] \label{thmexpclust}
There exist $C<\infty$ and $\mu>0$ such that
\begin{equation} \label{eqexpclust}
\E \left( \sup_{\psi, t} \mbox{Cor}_{\psi}(\tau_t(A_j), B_k)  \right) \le C \|A\| \|B\| e^{-\mu|j-k|}
\end{equation}
for all $L$, $1\le j,k \le L$, and all $A, B \in \C^{2\times 2}$. Here the supremum is taken over all normalized eigenstates $\psi$ of $H$ and all $t\in \R$.
\end{thm}

\vspace{.5cm}

Without going into details, we mention that \cite{SimsWarzel2016} has a related result for thermal states 
\begin{equation} 
\rho_{\beta} = e^{-\beta H_{XY}}/ \mbox{\rm{Tr}} e^{-\beta H_{XY}}.
\end{equation} 
where quantum expectations are defined as $\langle A \rangle_{\rho_{\beta}} = \mbox{\rm{Tr}} \rho_{\beta} A$ and one considers the quantum correlations $\langle \tau_t(A_j) B_k \rangle_{\rho_\beta} - \langle \tau_t(A_j) \rangle_{\rho_\beta} \langle B_k \rangle_{\rho_\beta}$.

Note that these results refer to {\it time-dependent} correlations, allowing for the observable $A_j$ to evolve in time. This is consistent with (and probably expected from) the LR bound \eqref{XYLR} which says that the dynamics does not significantly change the support of $A_j$.

We also mention that a much earlier related result was shown in \cite{KleinPerez1990} which only considers ground state correlations.

Next, the eigenstates of the disordered XY chain satisfy a uniform area law (in disorder average) at all energies. With the entanglement entropy defined as in \eqref{bipartite} and \eqref{entangle} above, one has
\begin{thm}[\cite{AbdulRahmanStolz2015}] \label{thmarealaw}
There exists $C<\infty$ such that
\begin{equation} \label{xyuniarealaw}
\E \left( \sup_{\psi} \mathcal{E}(\rho_{\psi}) \right) \le C 
\end{equation}
for all $L$ and all $1\le \ell < L$. Here the supremum is taken over all normalized eigenstates $\psi$ of $H_{XY}$.
\end{thm}

The method of proof of Theorem~\ref{thmarealaw} is essentially due to \cite{PasturSlavin2014}, who proved an area law only for the ground state of a disordered quasi-free Fermion system, but can handle the case of arbitrary dimension $d$ (where an area law means a bound $C\ell^{d-1}$ for a subsystem given by a $d$-dimensional cube of side length $\ell$). As discussed above, in $d=1$ this can be mapped to a result for the XY chain, but there is no similar correspondence between spin systems and free Fermion systems in dimension $d>1$.

Note that for the constant field isotropic XY chain ($\omega_j=h$ for all $j$ in \eqref{XYchain}), the large $\ell$ asymptotics of the {\it ground state} entanglement entropy can be found: In the subcritical case $|h|<2$, its leading term is $\frac{1}{3} \log_2 \ell$ (as proven in \cite{JinKorepin}, based on the formula \eqref{entropyformula} below), in the supercritical case $|h|>2$ the entanglement vanishes as the ground state is a product state. Thus the introduction of disorder eliminates the log-correction to the area law (and not just for the ground state).

We conclude this list with an {\it area law for the entanglement dynamics} under a quantum quench, which combines aspects of dynamical MBL and eigenstate MBL. Towards this, let
\begin{itemize}
\item $H_A$ and $H_B$ be the restrictions of $H$ to $A= [1,\ell]$ and $B=[\ell+1,L]$
\item $\psi_A$ and $\psi_B$ normalized eigenstates of $H_A$ and $H_B$, $\rho_A = |\psi_A\rangle \langle \psi_A|$, $\rho_B = |\psi_B\rangle \langle \psi_B|$
\item $\rho = \rho_A \otimes \rho_B$ the initial state at $t=0$ (so that  $\mathcal{E}(\rho)=0$)
\item $\rho_t = e^{-itH_{XY}} \rho e^{itH_{XY}}$ the dynamics under in the full spin system
\end{itemize}

This is viewed in physics as a quantum quench, where an interaction between the subsystems $A$ and $B$ is switched on for non-zero time. One asks how quickly an initially unentangled product state picks up entanglement through the interaction. For the disordered XY chain one gets the following dynamical area law, uniformly for all eigenstates of the subsystems, reflecting that in this fully many-body localized system the entanglement builds up only through effects near the surface between the subsystems.

\begin{thm}[\cite{ARNSS2016}] \label{thmdynent}
There exists $C<\infty$ such that
\[
\E \left( \sup_{t,\psi_A,\psi_B} \mathcal{E}(\rho_t) \right) \le C
\]
for all $\ell$ and $L$. Here the supremum is taken over all $t\in \R$ and all normalized eigenstates $\varphi_A$ and $\varphi_B$ of $H_A$ and $H_B$, respectively.
\end{thm}

In fact, a substantially more general result is proven in \cite{ARNSS2016}, which allows to choose the initial state $\rho$ as a more general product state $\rho = |\psi \rangle \langle \psi|$ with $\psi= \psi_1 \otimes \ldots \otimes \psi_m$ for eigenstates $\psi_k$ of the restriction of $H$ to intervals $\Lambda_k$, $k=1,\ldots,m$, for {\it any} decomposition $[1,L] = \Lambda_1 \cup \ldots \Lambda_m$. In particular, in the extreme case where each $\Lambda_k$ consists of only a single spin this shows that one can choose $\psi$ as an arbitrary up-down-spin product.

In the next section we will describe some of the properties of free Fermion systems which, together with proper ways of ``undoing'' the Jordan-Wigner transform \eqref{JordanWigner}, allow to reduce the proofs of Theorems~\ref{thmLR}, \ref{thmexpclust}, \ref{thmarealaw} and \ref{thmdynent} to Anderson localization of the effective Hamiltonian $M$ in the form \eqref{ecorloc}. This direct reduction of many-body localization to Anderson localization leads to the strongest results for the disordered XY chain. We mention, however, that there are also general results which establish a some degree of hierarchy between these different manifestations of MBL, in the sense that, for suitable classes of models, some of these properties imply versions of the others.

One such result shows that exponential clustering of eigenstates can be seen as a consequence of zero-velocity LR bounds: It was shown in \cite{HamzaSimsStolz2012} that exponential clustering of the ground state follows from a zero-velocity LR bound for the group velocity, without requiring a uniform ground state gap such as in the proofs of exponential clustering for the ground state in \cite{NachtergaeleSims2006} and \cite{HastingsKoma2006}. That similar reasoning is possible for excited states in systems with vanishing many-body transport is discussed in \cite{Friesdorfetal2015}.

Another general connection between MBL properties was established in \cite{BrandaoHorodecky2013, BrandaoHorodecky2015}, which gives a version of the fact that exponential clustering of eigenstates for a one-dimensional spin system implies that their bipartite entanglement satisfies an area law (thus, in particular, providing an alternative argument for Hastings' result \cite{Hastings} in the case of gapped systems).

Thus, in one-dimensional spin systems, we have some reasons to take guidance from the (slightly heuristic) hierarchy
\begin{equation} \label{hierarchy}
\mbox{Zero-Velocity LR} \quad \Longrightarrow \quad \mbox{Exponential Clustering} \quad \Longrightarrow \quad \mbox{Area Law}
\end{equation}
We don't know of any useful forms of converses of these statements (and don't believe that they exist).

\section{Comments on proofs and related results} \label{sec:XYproofs}

\subsection{Some ideas behind the proofs of Theorems~\ref{thmLR} to \ref{thmdynent}} As indicated, detailed proofs of Theorems~\ref{thmLR} to \ref{thmdynent} can be found in the papers \cite{HamzaSimsStolz2012}, \cite{SimsWarzel2016}, \cite{AbdulRahmanStolz2015} and \cite{ARNSS2016}. Here we will sketch some of the ideas behind the proofs, mostly focusing on how the reduction of MBL to Anderson localization is accomplished in these results.

The proof of Theorem~\ref{thmLR} in \cite{HamzaSimsStolz2012} proceeds in three steps: First, one uses the CAR to show that the Heisenberg dynamics of the Fermionic modes under the free Fermion system \eqref{freeFerm} is given by $\tau_t(b_j) = e^{-2it\lambda_j} b_j$. 

Second, it follows from this, $M= \mathcal{O}^T \Lambda \mathcal{O}$ and $b= \mathcal{O}c$ that 
\begin{equation} 
\tau_t(c_j) = \sum_{\ell} \left( e^{-2iMt} \right)_{j\ell} c_{\ell},
\end{equation}
and it is through this that dynamical Anderson localization \eqref{dynloc} is exploited in the proof. 

From this, in the third step, one can determine the dynamics of the lowering operators $a_j$ by inverting the Jordan-Wigner transform,
\begin{equation}
a_j = \sigma_1^z \ldots \sigma_{j-1}^z c_j = (2c_1^* c_1-\idty) \ldots (2c_{j-1}^* c_{j-1} -\idty) c_j.
\end{equation}
In concrete terms this requires iterative applications of the Leibnitz rule $[AB,C] = A[B,C] + [A,C]B$. After summing several geometric series this leads to a proof of \eqref{XYLR} if $A_j$ is $a_j$, $a_j^*$, $a_j a_j^*$ or $a_j^* a_j$ and the corresponding four choices for $B_k$. As these operators form a basis of the one-site observables, the proof is complete.

The main tool in the proofs of Theorems~\ref{thmexpclust}, \ref{thmarealaw} and \ref{thmdynent} is the theory of quasifree states and, in particular, the fact that the latter are entirely characterized by their two-point correlation matrices. We only recall a few key facts here:

\begin{itemize}

\item All eigenstates $\rho=\rho_{\alpha} = |\varphi_\alpha \rangle \langle \varphi_\alpha|$, $\alpha\in \{0,1\}^L$ (see \eqref{varphialpha}), and thermal states $\rho= \rho_{\beta} = e^{-\beta H}/\tr e^{-\beta H}$, $0<\beta <\infty$, of a quasifree Fermion system $H=c^*M c$ are quasifree. This means that expectations of arbitrary products of the $c_j$ and $c_j^*$ can be calculated by Wick's Rule in terms of the correlation matrix
\begin{equation}
\Gamma_{\rho} = \left( \langle c_j c_k^* \rangle_{\rho} \right)_{j,k=1}^L.
\end{equation}
In more mathematical terms this means the evaluation of Pfaffians of the latter.

\item For the eigenstates $\rho_\alpha$ we can express the correlation matrix through spectral projections of the effective Hamiltonian $M$: If $\sigma(M) = \{\lambda_j:j=1,\ldots,L\}$ is simple (which holds almost surely), then 
\begin{equation}  \label{Gammaalpha}
\Gamma_{\rho_\alpha} = \chi_{\Delta_\alpha}(M) = \sum_{\ell:\lambda_\ell \in \Delta_\alpha} |\varphi_\ell \rangle \langle \varphi_\ell |, 
\end{equation} 
where $\Delta_\alpha := \{\lambda_j: \alpha_j=1\}$. Note that as a special case of \eqref{ecorloc} we have
\begin{equation} \label{projectorloc}
\E \left( \sup_{\alpha} |(\chi_{\Delta_\alpha}(M))_{jk}| \right) \le C e^{-\mu|j-k|}.
\end{equation}

\item The calculation of correlations of local observables in Theorem~\ref{thmexpclust} can now be reduced to evaluating Pfaffians of matrixes with off-diagonal exponential decay (on average) given by \eqref{projectorloc}. To see that the result still has the exponential decay \eqref{eqexpclust} requires a suitable strategy for row and column elimination in the Pfaffian provided in \cite{SimsWarzel2016} (a naive application of H\"older's inequality to factorize the expectation of the appearing products would lead to collapse of the decay rate $\mu$).

\item Correlation matrices can also be used to show the entanglement bound in Theorem~\ref{thmarealaw}. In our setting, where $A=[1,\ell]$, $B=[\ell+1,L]$ and the Fermionic modes are given by the ``left-local'' Jordan-Wigner transform \eqref{JordanWigner}, the reduced state $\rho^A$ of a quasifree state $\rho$ is again quasifree and its correlation matrix is the upper $\ell \times \ell$-block of $\Gamma_{\rho}$,
\begin{equation}
\Gamma_{\rho^A} = \left( \langle c_j c_k^* \rangle_{\rho} \right)_{j,k=1}^\ell.
\end{equation}

\item By a fact which can be found in \cite{VLRK2003} (but most likely was known to physicists before), the entanglement entropy of a quasifree state $\rho$ can now be reduced to
\begin{equation} \label{entropyformula}
\mathcal{E}(\rho) = \mathcal{S}(\rho^A) = -\mbox{\rm{Tr}}\,\rho^A \log \rho^A = -\mbox{\rm{tr}} \,h(\Gamma_{\rho^A}),
\end{equation}
where $h(x)=x\log x  +(1-x) \log(1-x)$. Note here that Tr denotes the trace in the $2^{\ell}$-dimensional space $\mathcal{H}_\ell$ and tr is the trace in the $\ell$-dimensional space $\C^{\ell}$. From here an argument first provided for the ground state of a quasi-free Fermion system in \cite{PasturSlavin2014} leads to the area law \eqref{xyuniarealaw}.
\item Concerning Theorem~\ref{thmdynent} we mention three more facts, referring to \cite{ARNSS2016} for more careful formulations and proofs: (i) A quasi-free state with respect to the Fermionic modes $c_j$ remains quasi-free under the Heisenberg dynamics of \eqref{quad} (Lemma~5.1 in \cite{ARNSS2016}). (ii) Tensor products of quasi-free states are quasi-free (Lemma~5.2 in \cite{ARNSS2016}). (iii) Based on (i) and (ii) the proof of Theorem~\ref{thmdynent} reduces to an analysis of correlation matrices and the fact that all three operators $H$, $H_A$ and $H_B$ have effective Hamiltonians $M$, $M_A$ and $M_B$ which are localized in the sense \eqref{ecorloc}.

\end{itemize}

\subsection{Extensions to the Anisotropic XY chain} \label{sec:anisoXY}

We conclude our discussion of the disordered XY chain by mentioning that much of what was done above can be extended to the anisotropic version of the model,
\begin{eqnarray} \label{anisoXY}
H_{\gamma} & = & -\sum_{j=1}^{L-1} ((1+\gamma)\sigma_j^X \sigma_{j+1}^X + (1-\gamma) \sigma_j^Y \sigma_{j+1}^Y) - \lambda \sum_{j=1}^L \omega_j \sigma_j^Z \\
& = & \mathcal{C}^* \tilde{M} \mathcal{C} + E_0 \idty \notag,
\end{eqnarray}
where $\gamma \not=0$ introduces an anisotropy between the X and Y parts of the interaction. Here, in the second line we use the operator-valued column and row vectors $\mathcal{C} = (c_1,\ldots,c_L,c_1^*,\ldots,c_L^*)^t$ and $\mathcal{C}^* = (c_1^*,\ldots,c_L^*,c_1,\ldots,c_L)$, with the Jordan-Wigner Fermionic modes $c_j$ given again by \eqref{JordanWigner}. The effective Hamiltonian is now the $2L \times 2L$ block matrix 
\begin{equation} \label{blockAnderson}
\tilde{M} = \begin{pmatrix} M & K \\ -K & -M \end{pmatrix}
\end{equation}
with
\begin{equation} 
M = \begin{pmatrix} \omega_1 & -1 & & \\ -1 & \ddots & \ddots & \\ & \ddots & \ddots & -1 \\ & & -1 & \omega_L \end{pmatrix}, \quad K = \begin{pmatrix} 0 & -\gamma & & \\ \gamma & \ddots & \ddots & \\ & \ddots & \ddots & -\gamma \\ & & \gamma & 0 \end{pmatrix}.
\end{equation}
This means that in the disordered case the effective Hamiltonian couples two copies of the Anderson model (one positive and one negative) via the off-diagonal block $K$, the latter implementing the anisotropy. Much of what has been said about the isotropic disordered XY chain in Section~\ref{XYsurvey} can be extended to the anisotropic model as long as one can show an analogue of the eigencorrelator localization property \eqref{ecorloc} for the {\it block Anderson model} \eqref{blockAnderson}. This has been done if the disorder parameter $\lambda$ introduced in \eqref{anisoXY} is sufficiently large, where it follows as a special case of a result in \cite{ElgartShamisSodin}. As a consequence, in the large disorder regime Theorems~\ref{thmLR} to \ref{thmdynent} extend to the anisotropic case, as discussed in the papers cited above. 

The situation is less clear for small disorder, where one still might expect that $\tilde{M}$ is fully localized, due to the one-dimensionality of the model. But it hasn't been fully understood if the block Hamiltonian $\tilde{M}$ may have a zero-energy singularity (which would then affect the full many-body spectrum through \eqref{varphialpha}), see \cite{ChapmanStolz} for more discussion.

\section{Interlude: The Ising Model} \label{sec:Ising}

As was confirmed in our discussion up to this point, the XY chain is a substantially simpler model than the classical Heisenberg model $\sum_j (\sigma_j^X \sigma_{j+1}^X + \sigma_j^Y \sigma_{j+1}^Y + \sigma_j^Z \sigma_{j+1}^Z)$, which features an extra interaction term between neighboring spins in terms of $Z$ Pauli matrices. Trying to understand MBL properties of disordered versions of the full Heisenberg model is indeed a much harder and still widely open problem. In fact, it is believed in physics and supported by substantial numerical evidence (e.g.\ \cite{Znidaricetal, PalHuse}) that the Heisenberg chain in random field will give rise to a many-body localization-delocalization transition. Thus, as opposed to the fully many-body localized disordered XY chain, we can expect MBL phenomena for the Heisenberg chain to only appear at sufficiently low energy or in the case of strong disorder. A survey of some first rigorous results of this kind will be the topic of the remainder of our presentation. 

In some sense, we will exploit phenomena at the opposite extreme of what we did for the XY chain. In the study of the latter it was crucial that no Z-term was included in the interaction. Now we will look at an anisotropic version of the Heisenberg model, where the Z-term dominates the XY-terms, the so-called Ising phase of the XXZ chain.

To understand what to expect it is instructive to start with the Ising limit of the XXZ chain, where the XY-part of the interaction is dropped altogether. We will do this in infinite volume, where a disordered version of this model is given by
\begin{equation} \label{Ising}
H_{Ising}(\omega) = \frac{1}{4} \sum_{j\in \Z}(I-\sigma_j^Z \sigma_{j+1}^Z) + \sum_{j\in \Z} \omega_j {\mathcal N}_j.
\end{equation}
Here the disorder is introduced in the form of random couplings at the {\it local number operators} 
\begin{equation} 
\mathcal{N}_j= \begin{pmatrix} 0 & 0 \\ 0 & 1 \end{pmatrix}_j,
\end{equation} 
which count the number of down spins. Note that $\mathcal{N}_j=\frac{1}{2}(I-\sigma_j^Z )$, so that the random field in \eqref{Ising} differs from the field in the XY chain \eqref{XYchain} only by an energy shift. For the distribution of the random parameters we will again assume \eqref{distribution} and now in addition that
\begin{equation} \label{posrandom}
\supp \rho = [0,\omega_{max}] \:\: \mbox{for some $\omega_{max}>0$}.
\end{equation}

The model $H_{Ising}(\omega)$ is trivial in the sense that it is diagonal in the product basis
\begin{equation} 
\varphi_X:= \prod_{j\in X} a_j^* \Omega, \quad X\subset \Z \:\,\mbox{finite},
\end{equation}
with $\Omega = |\ldots \uparrow \uparrow \uparrow \ldots \rangle$ the all up-spins vector. By ``choosing'' the set $\{\varphi_X: X\subset \Z \;\mbox{finite}\}$ as an ONB of the Hilbert space (which is often referred to as the {\it incomplete tensor product} of infinitely many copies of $\C^2$), we have made a choice consistent with a particular implementation of the thermodynamic limit via the GNS construction. Here $\Omega =\varphi_{\emptyset}$ takes the role of the GNS vaccum vector. The operators 
\begin{equation} 
a_j^* = \begin{pmatrix} 0 & 0 \\ 1 & 0 \end{pmatrix}_j
\end{equation}
take the role of creation operators of a down-spin (``particle'') at site $j$. Thus all basis states have finitely many down-spins in a sea of up-spins. All calculations in the remainder of this work are made by identifying
\begin{equation} 
|\hspace{-.1cm}\uparrow\rangle = \begin{pmatrix} 1 \\ 0 \end{pmatrix} \quad \mbox{and} \quad |\hspace{-.1cm} \downarrow\rangle = \begin{pmatrix} 0 \\ 1 \end{pmatrix}.
\end{equation}

One checks that
\begin{equation}
(I-\sigma_j^Z \sigma_{j+1}^Z) \varphi_X = \left\{ \begin{array}{ll} 2\varphi_X, & \mbox{if} \:\{j,j+1\} \in \partial X,  \\ 0, & \mbox{else.} \end{array} \right.
\end{equation}
Here we have written
\begin{equation} \label{surface}
\partial X := \{(j,j+1):\:(j,j+1) \cap X \not= \emptyset, \: (j,j+1) \cap X^c \not= \emptyset \}
\end{equation}
for the graph theoretic surface of $X$ in $\Z$ (which in our 1D setting is twice the number of connected components of $X$). From this it follows that
\begin{equation} \label{Isingenergies}
H_{Ising}(\omega) \varphi_X = \Big( \frac{1}{2} |\partial X| + \sum_{j\in X} \omega_j \Big) \varphi_X
\end{equation}
and thus we have explicitly diagonalized the model.

For $\omega=0$, we have that the vacuum $\Omega$ is the unique ground state of $H_{Ising}(0)$ to $E_0=0$. The other eigenvalues are $E_k=k \in \{1,2,3,\ldots\}$, and for each $k$ the eigenspace is spanned by those $\varphi_X$ for which $X$ has $k$ connected components, i.e., the down spins form $k$ clusters. In particular, eigenstates to the lowest non-zero eigenvalue $E_1=1$ consist of a single droplet of down-spins in a sea of up-spins, but these droplets can have arbitrary size and position, leading to infinite degeneracy (as is also the case for the higher energies).

There are two reasons for having included this discussion of the trivial Ising chain here. First, as we will see in the next section, the droplet structure of low energy eigenstates also appears in the so-called Ising phase of the XXZ model.

The second reason is more philosophical, in that one can ask if the deterministic Hamiltonian $H_{Ising}(0)$ should be considered as many-body localized. After all, the system has a full set of eigenstates given by product states, all having trivial spatial correlations and vanishing entanglement in the sense of Theorems~\ref{ExpCl} and \ref{AreaLaw}. However, the eigenspaces of $H_{Ising}(0)$ are highly degenerate, in fact, infinitely degenerate in the infinite volume case we have considered here. Suitable linear combinations of the states $\varphi_X$ can be found which give eigenstates with long range correlations and high entanglement. For example, consider the normalized linear combination 
\begin{equation} \label{satexample}
\varphi =\ell^{-1/2} \sum_{j=1}^\ell \varphi_{[j,j+\ell-1]}
\end{equation} 
of single droplet states $\varphi_{[j,j+\ell-1]}$, so that $\varphi$ is an eigenvector to $k=1$. A simple calculation shows that its entanglement with respect to the subsystem $[1,\ell]$ grows like $\log \ell$, thus violating the strict area law we expect for an MBL regime.

Second, being in the localized regime of a quantum many-body Hamiltonian should not be an ``accidental'' fact, meaning that there should be some stability of the associated phenomena under perturbations. This should include the possibility of spatially extensive perturbations, e.g.\ small perturbations of the next-neighbor interaction term $h_{j,j+1}$ in a Hamiltonian of the form \eqref{Ham}, uniformly at all sites $j$. In our discussion of the XXZ chain below we will see that this is not the case for the zero-field Ising chain: In the absence of the random field term $\sum_j \omega_j \mathcal{N}_j$ in \eqref{Ising}, the addition of an arbitrarily small XX-term to the next-neighbor interaction in the Ising chain will radically change the structure of the eigenstates of the model, putting the model out of the MBL phase.

On the other hand, the addition of the random field in $H_{Ising}(\omega)$ can be seen as remedying  these shortcomings and putting the model more firmly into the realm of MBL. The addition of the field energy in \eqref{Isingenergies} will diffuse the high degeneracies of the spectrum, thus excluding the possibility of building highly correlated or entangled states by linearly combining product states within a fixed eigenspace. In fact, when restricted to finite volume the Ising chain in random field almost surely has non-degenerate spectrum (by an argument in Appendix~1 of \cite{AbdulRahmanStolz2015} which also applies to the Ising chain). Thus, almost surely, the trivially correlated and entangled product states $\varphi_X$ will be the {\it only} eigenstates.

More convincing support of our point that the MBL regime is only created after the introduction of disorder into the model may come from the results for the less trivial model of the disordered XXZ chain to be presented below. When starting from the disordered Ising chain, these results can be interpreted as stability of the MBL regime under a sufficiently small change of the interaction terms in the Hamiltonian (and at sufficiently low energies).

\section{The Droplet Spectrum in the XXZ Chain} \label{sec:XXZdet}

In this and the following section we present recent results on many-body localization properties in the droplet regime of the disordered XXZ chain. We start in the current section by describing this model in the deterministic setting. The droplet regime and its special features were first pointed out and studied  in \cite{Starr2001, NachtergaeleStarr2001}, see also \cite{NachtergaeleSpitzerStarr2007}. This topic was then revisited in the MS thesis \cite{Fischbacher2013} and the related publication \cite{FischbacherStolz2013} (as well as in \cite{FischbacherStolz2017} for more recent extensions) from the point of view of laying ground work to study the disordered version of the model. This disordered XXZ chain and, in particular, its droplet regime had been proposed by B.\ Nachtergaele as a likely candidate for the presence of MBL features. That this can indeed be proven rigorously, in essentially all its expected manifestations, was recently demonstrated in the papers \cite{BeaudWarzel2017a}, \cite{EKS2017a}, \cite{EKS2017b} and \cite{BeaudWarzel2017b}, as will be discussed in Section~\ref{MBLXXZ} below.

\subsection{The XXZ chain in external field}

We model the disordered infinite XXZ chain by the Hamiltonian
\begin{equation} \label{randomXXZ}
 H_{XXZ}(\omega) = \sum_{j\in \Z} h_{j,j+1} + \sum_{j\in \Z} \omega_j {\mathcal N}_j.
 \end{equation}
Here, as for the Ising chain above, in the proper form of the thermodynamic limit this is considered as an operator on the Hilbert space completion of the orthonormal system $\{\varphi_X: X \subset \Z \:\mbox{finite}\}$, i.e., the up-down-spin product states with finitely many down-spins at the sites $X$. For the next-neighbor interaction we choose
\begin{eqnarray} \label{hjjdef}
 h_{j,j+1} & = & \frac{1}{4}(\idty-\sigma_j^Z \sigma_{j+1}^Z) - \frac{1}{4\Delta} (\sigma_j^X \sigma_{j+1}^X + \sigma_j^Y \sigma_{j+1}^Y) \\
 & = & \frac{1}{4\Delta} (\idty - \vec{\sigma}_j \cdot \vec{\sigma}_{j+1}) + \frac{1}{4} \left(\idty-\frac{1}{\Delta}\right) (\idty-\sigma_j^Z \sigma_{j+1}^Z) \notag
 \end{eqnarray}

We will always assume $\Delta>1$, characterizing the  ``Ising phase" of the XXZ model, where the Ising part of the interaction dominates the XY part. At this point the parameters $\omega_j$ are deterministic and we will merely assume their non-negativity. That they represent random variables will only become relevant in Section~\ref{MBLXXZ}.

A key property of the XXZ chain is that it preserves the particle number, i.e., that the $N$-particle ($N$-down-spin, $N$-magnon) subspaces
\begin{equation} 
{\mathcal H}_N = \overline{\mbox{span}\{\varphi_X: |X| = N\}}, \:\: N=0,1,2,\ldots
\end{equation}
are invariant under $H_{XXZ}(\omega)$. Thus we can decompose
\begin{equation} 
H_{XXZ}(\omega) = \bigoplus_{N=0}^\infty H_N(\omega).
\end{equation}
${\mathcal H}_0$ is the one-dimensional space spanned by the all up-spins vacuum vector $\Omega$ and $H_0(\omega)\Omega =0$, making $\Omega$ the non-degenerate ground state of $H_{XXZ}(\omega)$ (as will become clear from the following).
To further describe the $N$-particle operators $H_N(\omega)$ for $N\ge 1$, we unitarily identify 
${\mathcal H}_N \cong \ell^2({\mathcal V}_N)$, where
\begin{equation} 
{\mathcal V}_N = \{x=(x_1,\ldots,x_N)\in \Z^N: x_1< x_2 < \ldots < x_N\}, 
\end{equation}
i.e., $x$ is the ordered labeling of the down spin configuration $X=\{x_1,\ldots,x_N\} \subset \Z$, so that $\varphi_X$ corresponds to the canonical basis vector $\delta_x$. In this identification, sometimes called the {\it hard core particle formulation} of the XXZ chain,
\begin{equation} \label{HN}
H_N(\omega) = -\frac{1}{2\Delta} A_N + \frac{1}{2}  D_N +  V_{\omega} \quad \mbox{on $\ell^2({\mathcal V}_N)$} .
\end{equation}
Here

\begin{itemize}

\item $A_N$ is the adjacency operator on ${\mathcal V}_N$,
\begin{equation}
(A_N f)(x) = \sum_{y \sim x} f(y),
\end{equation}
where $y\sim x$ denotes a next neighbor in the $\ell^1$-distance on $\mathcal{V}_N$ inherited from $\Z^N$.

\item $D_N$ is a multiplication operator (``potential'') on $\ell^2({\mathcal V}_N)$ by
\begin{eqnarray} 
D_N(x) = 2\left|\{\mbox{connected components of $X$}\}\right| = |\partial X|,
\end{eqnarray}
compare \eqref{surface}. Another useful way to understand $D_N$ is as the degree function on the graph ${\mathcal V}_N$, i.e., $D_N(x)$ is the number of next neighbors of $x$ in the graph ${\mathcal V}_N$.

\item $V_\omega$ is an $N$-particle potential, 
\begin{equation} 
V_{\omega}(x) = \omega_{x_1} + \ldots + \omega_{x_N}.
\end{equation}

\item In order to best understand the positivity properties of $H_N(\omega)$ it is helpful to use that $A_N= L_N+D_N$, where $L_N\le 0$ is the graph Laplacian on ${\mathcal V}_N$,
\begin{equation}
(L_N f)(x) = \sum_{y\sim x} (f(y)-f(x)).
\end{equation}
This yields
\begin{equation} \label{HNalt}
H_N(\omega) = -\frac{1}{2\Delta} L_N + \frac{1}{2}\left(1-\frac{1}{\Delta}\right)  D_N +  V_{\omega},
\end{equation}
where by our assumptions all three terms are non-negative.

\end{itemize}

\subsection{The droplet regime of the free XXZ chain}

We start by collecting some facts for the case $\omega=0$, i.e., for the {\it free XXZ chain}
\begin{equation} 
H_{XXZ}(0) = \bigoplus_{N\ge 0} H_N(0).
\end{equation}

The free infinite volume XXZ chain is a rare example of a quantum many-body system which can be exactly diagonalized by the Bethe ansatz, meaning that for each $N$ one can explicitly determine the spectrum of $H_N(0)$ as well as a complete set of generalized eigenvectors, see \cite{BabbittGutkin, Borodinetal} for two different approaches to this. But we also refer to \cite{FischbacherStolz2013} (drawing on \cite{NachtergaeleStarr2001} and \cite{NachtergaeleSpitzerStarr2007}), where all that is relevant to the discussion here has been re-derived by more elementary means. 

We summarize relevant fact in several remarks:

\begin{itemize}

\item $H_0(0)$ on the one-dimensional subspace spanned by $\Omega$ provides the non-degenerate ground state energy $E_0=0$ for $H_{XXZ}(0)$. For all other $N$ one has $D_N\ge 2$ and thus, by \eqref{HNalt}, $H_N(0) \ge 1- \frac{1}{\Delta}$, yielding a ground state gap for the free infinite XXZ chain in the Ising phase.

\item For $N\ge 1$ the spectrum of $H_N(0)$ below $2(1-\frac{1}{\Delta})$ is given by
\begin{equation}
\sigma(H_N(0)) \cap \left[1-\frac{1}{\Delta}, 2\left(1-\frac{1}{\Delta}\right) \right] = \delta_N \cap \left[1-\frac{1}{\Delta}, 2\left(1-\frac{1}{\Delta}\right) \right],
\end{equation}
where
\begin{equation}
\delta_N  =  \left[ \tanh(\rho) \cdot \frac{\cosh(N\rho)-1}{\sinh(N\rho)}, \tanh(\rho) \cdot \frac{\cosh(N\rho)+1}{\sinh(N\rho)}\right] 
\end{equation}
with $\rho>0$ determined by $\cosh(\rho) = \Delta$. The intervals $\delta_N$ are nested as $N$ increases and tend to the single point $\sqrt{1-1/\Delta^2}$ as $N\to\infty$. The first few are given by
\begin{equation}
\delta_1 = \left[ 1-\frac{1}{\Delta}, 1+ \frac{1}{\Delta} \right], \quad \delta_2 = \left[ 1-\frac{1}{\Delta^2}, 1 \right], \quad \delta_3 = \left[ 1- \frac{1}{2\Delta^2-\Delta}, 1 - \frac{1}{2\Delta^2+\Delta} \right].
\end{equation}

\item Exact formulas for the eigenstates to energies in the bands $\delta_N$ can be given, see \cite{NachtergaeleStarr2001}. These, up to exponentially small corrections (with a rate increasing in $\Delta$), have the form of linear combinations of single clusters of down-spins (i.e., the droplets which give the exact eigenstates of the Ising chain discussed in Section~\ref{sec:Ising}). 

A qualitative way to express this, which also holds in finite volume and under addition of a field $\omega_j \ge 0$, will be provided in Theorem~\ref{thm:evdecay} below. On a heuristic level one can understand this from the hard core particles formulation of the XXZ chain: $H_N(0)  =  - \frac{1}{2\Delta} A_N + \frac{1}{2} D_N$, where the potential is attractive: 
\begin{equation} 
D_N(x) = N- \sum_{1\le k < \ell \le N} Q(|x_k-x_{\ell}|)
\end{equation} 
with $Q(1)=1$, $Q(r)=0$ for $r\not= 1$. Thus the energy should be minimized if the $N$ particles occupy neighboring sites (form a droplet), at least for weak hopping, which will turn out to mean $\Delta>1$.

\end{itemize}

These observations lead us to refer to $\delta_N$ as the droplet band of $H_N(0)$ and to
\begin{equation} \label{dropspec}
I = \left[ 1-\frac{1}{\Delta},2\Big(1-\frac{1}{\Delta}\Big)\right)  
\end{equation}
as the {\it droplet spectrum} of $H_{XXZ}(0)$. Note, however, that for $1<\Delta<3$ and $N=1$ part of the droplet band extends above $2(1-\frac{1}{\Delta})$. Also, for $\Delta>3$ there is a gap between $\cup_{N\ge 1} \delta_N$ and the higher spectral bands. It is not hard too see by a Weyl sequence argument, compare \cite{FischbacherStolz2013}, that this gap will be filled after adding a random field. In fact, the almost sure spectrum of $H_{XXZ}(\omega)$, under the assumptions  \eqref{distribution} and \eqref{posrandom}, is $\{0\} \cap [1-\frac{1}{\Delta},\infty)$. The remainder of our discussion can be understood as showing that the introduction of disorder leads to many-body localization of the droplet spectrum $I$.

\subsection{The finite volume XXZ chain with droplet boundary conditions}

The remainder of our discussion refers to finite volume restrictions of the XXZ chain. More precisely, for arbitrary $L\in \N$ we work in the Hilbert space ${\mathcal H}_{[-L,L]} = \bigoplus_{j=-L}^L \C^2$ with Hamiltonian
\begin{equation} \label{finiteXXZ}
H_{XXZ}^L(\omega) = \sum_{j=-L}^{L-1} h_{j,j+1} +\sum_{j=-L}^L \omega_j {\mathcal N}_j + \beta ({\mathcal N}_{-L} + {\mathcal N}_L).
\end{equation}
Here we have to introduce the so-called {\it droplet boundary condition} $\beta ({\mathcal N}_{-L} + {\mathcal N}_L)$ (e.g.\ \cite{NachtergaeleStarr2001}) with 
\begin{equation} \label{beta} 
\beta \ge \frac{1}{2} (1-\frac{1}{\Delta})
\end{equation} 
 to make up for the lost interaction terms at the boundary and preserve crucial properties such as the positivity in \eqref{HNalt}, as we will see below. While working in finite volume, we stress that all our results below will provide bounds which hold uniformly in the system size $L$.

As in infinite volume, we still have particle number conservation:
\begin{equation} 
H_{XXZ}^L(\omega) = 0 \oplus \bigoplus_{N=1}^{2L+1} H_N^L(\omega)\: \:\mbox{on \:\:span} \{\Omega\}  \oplus \bigoplus_{N=1}^{2L+1} \ell^2({\mathcal V}_N^L),
\end{equation}
where $\Omega = \otimes_{j=-L}^L\: |\hspace{-.1cm}\uparrow \rangle$ now denotes the finite volume vacuum vector and
\begin{equation} 
{\mathcal V}_N^L := \{ x=(x_1,\ldots,x_N) \in \Z^N: -L \le x_1 < \ldots < x_N \le L\}.
\end{equation}
The finite volume restriction of the N-particle operator is
\begin{equation} \label{HNL}
H_N^L(\omega) = -\frac{1}{2\Delta} L_N + \frac{1}{2}\left(1-\frac{1}{\Delta}\right)  D_N + V_{\omega} + \left(\beta -\frac{1}{2}\big(1-\frac{1}{\Delta}\big)\right) \chi^{(L)}.
\end{equation}
Here we have included a boundary term, where $\chi^{(L)} = \chi_L+\chi_{-L}$, with $ \chi_L$ and $\chi_{-L}$ denoting the characteristic functions of the right and left boundaries $\{(x_1,\ldots,x_N): x_N=L\}$ and $\{(x_1,\ldots,x_N): x_1=-L\}$ of $\mathcal{V}_N^L$. All four terms on the right of  \eqref{HNL} are non-negative, highlighting the various positivity properties of our model which enter the proofs of the following results in many ways. In particular, this explains the assumption \eqref{beta} on $\beta$ made above.

We will abbreviate $H=H_{XXZ}^L(\omega)$ and $H_N=H_N^L(\omega)$ for the rest of our presentation. 

We point out that, as opposed to the infinite volume XXZ chain, the finite volume model is {\it not} exactly solvable, not even for $\omega=0$. However, all that is needed for the following discussion of localization properties can be shown by qualitative arguments. In particular, the appearance of a droplet regime for energies below $2(1-\frac{1}{\Delta})$ is reflected in the following result on the decay properties of eigenvectors.

For this let
\begin{equation} 
{\mathcal V}_{N,1}^L := \left\{ x= (x_1, x_1+1, \ldots, x_1+N-1) \in {\mathcal V}_N^L\right\} 
\end{equation}
denote the droplet configurations in ${\mathcal V}_N^L$, i.e., configurations in which all $N$ particles occupy neighboring sites. 

In all the results discussed below, we will need a small ``safety distance'' from the upper end of the droplet spectrum \eqref{dropspec}. Thus, for any $\delta>0$, we write
\begin{equation}
I_\delta := \left[1-\frac{1}{\Delta}, (2-\delta)\Big(1-\frac{1}{\Delta}\Big)\right].
\end{equation}

\begin{thm}[Droplet structure of low energy states] \label{thm:evdecay}
Let $\Delta>1$ and $\delta>0$. Then there exist constants $C=C(\delta)$ and $\mu = \mu(\Delta,\delta)>0$, such that for every $\omega\ge 0$, every normalized vector $\psi$ in the range of $P_{I_\delta}(H_N)$ (the spectral projection of $H_N$ onto $I_\delta$), and every set ${\mathcal A} \subset {\mathcal V}_N^L$, it holds that
\begin{equation} \label{evdecay} 
\| \chi_{\mathcal A} \psi \| \le C e^{-\mu d_N({\mathcal A}, {\mathcal V}_{N,1}^L)} \| \chi_{{\mathcal V}_{N,1}^L} \psi \|. 
\end{equation}
\end{thm}

Here 
\begin{equation} 
d_N(\mathcal{A}, \mathcal{B}) = \min_{x\in \mathcal{A}, y \in \mathcal{B}} \sum_{j=1}^N |x_j-y_j|
\end{equation} 
denotes the $\ell^1$-distance of two subsets of $\mathcal{A}$, $\mathcal{B}$ of ${\mathcal V}_N^L$.

 The bound \eqref{evdecay} is a qualitative version of the explicit formulas in infinite volume from \cite{NachtergaeleStarr2001}, describing the droplet structure of eigenvectors to energies below the threshold $2(1-1/\Delta)$. It says that, up to exponentially decaying corrections, the mass of all these eigenvectors is concentrated at the droplet configurations $\mathcal{V}_{N,1}^L$ (but not necessarily near a single one of these configurations). 

The proof of Theorem~\ref{thm:evdecay} follows from a Combes-Thomas type bound.  An important feature of the bound \eqref{evdecay}, as well as of the underlying Combes-Thomas bound, is that the constants can be chosen uniform not only in the system size $L$, but also in the particle number $N$ (and thus the dimension of $\mathcal{V}_N$). The standard proof of a Combes-Thomas bound (e.g.\ \cite{Kirsch}) would lead to a $1/N$-dependence on the rate of exponential decay on the right hand side of \eqref{evdecay}. That this can be avoided in our case is due to the fact that in the Schr\"odinger-type operator \eqref{HN} the hopping part $-A_N/(2\Delta)$ is ``balanced'' by the potential part $D_N/2$, in the sense that $D_N$ is exactly the number of hoppings originating at any site $x$ (and that the potential part dominates the hopping part due to $\Delta>1$). 

Combes-Thomas bounds of this form do not just readily imply Theorem~\ref{thm:evdecay}, but are the key fact behind all the localization properties of the droplet spectrum of the random XXZ chain which we will present in Section~\ref{MBLXXZ}.
Due to its central importance, we will discuss one of the incarnations of this Combes-Thomas bound, as well as the main ideas behind its proof and how it implies Theorem~\ref{thm:evdecay}, in some more detail in Section~\ref{sec:CT} below.

As mentioned, \eqref{evdecay} is a deterministic result, holding uniformly for any choice of the disorder parameters $\omega \ge 0$ and, in particular, for $\omega=0$. This is {\it almost} a localization property of these states, up to the fact that the position of the droplet (labeled by, say, its left endpoint $x_1$) remains as a single one-dimensional spatial degree of freedom. It is in this sense that the droplet states of the XXZ chain can be viewed as describing a single one-dimensional {\it quasi-particle} with position quantized by $x_1$. 

In this view, the following results on many-body localization are interpreted as saying that the addition of disorder localizes this remaining degree of freedom. One can therefore use the proof of localization for the one-dimensional Anderson model as a heuristic guideline for the latter proofs of localization properties of the disordered XXZ chain (where the detailed implementation of the heurisitics requires to overcome a number of substantial technical challenges).

\section{MBL properties of the XXZ chain} \label{MBLXXZ}

We now survey the MBL properties of the droplet spectrum of the disordered XXZ chain $H:=H_{XXZ}^L(\omega)$ \eqref{finiteXXZ}  which have been proven in the works \cite{BeaudWarzel2017a,BeaudWarzel2017b,EKS2017a,EKS2017b}. Throughout this section we refer to $H_{XXZ}^L(\omega)$ as defined through \eqref{randomXXZ}, \eqref{hjjdef} and \eqref{finiteXXZ}, with $\Delta>1$, $\beta\ge (1-\frac{1}{\Delta})/2$ and assumptions on the disorder given by \eqref{distribution} and \eqref{posrandom}.

We start with entanglement bounds, where we consider any bipartite decomposition: $[-L,L] = \Lambda_{\ell} \cup \Lambda_r$ of the chain into two intervals. As above we write $\mathcal{E}(\rho_{\psi}) := \mathcal{S}(\rho_{\psi}^{\Lambda_\ell})$ for the entanglement entropy of a normalized state $\psi$ with respect to this decomposition.

\begin{thm}[\cite{BeaudWarzel2017b}] \label{thm:XXZarealaw}
Let $\Delta>1$ and $\delta>0$. Then there exist finite constants $C_1 = C_1(\Delta,\delta)$ and $C_2 = C_2(\Delta,\delta)$ such that

(a) for every $\omega\ge 0$ and every normalized vector $\psi \in P_{I_\delta}(H)$,
\begin{equation} 
{\mathcal E}(\rho_{\psi}) \le C_1 \log |\Lambda_{\ell}|.
\end{equation}

(b) If $\E(\cdot)$ denotes disorder averaging, then
\begin{equation} 
\E \Big( \sup_{\psi}  {\mathcal E}(\rho_{\psi_E}) \Big) \le C_2,
\end{equation}
where the supremum is taken over all normalized vectors $\psi$ in the range of $P_{I_\delta}(H)$.
\end{thm}

We make several remarks about this result and the strategy for its proof in \cite{BeaudWarzel2017b}:

\begin{itemize}

\item Part (a) is entirely deterministic. In particular, it holds for $\omega=0$. It provides a log-corrected version of a one-dimensional area law. This means that low energy states, even without any localizing effects due to disorder, have quite low entanglement. However, by itself this can not be seen as an MBL property, as such log-corrected area laws are also known for a number of examples of deterministic non-gapped spin chain models (compare, for example, the results for the deterministic XY chain mentioned after Theorem~\ref{thmarealaw} above). 

\item Note the the $\log \ell$ upper bound is saturated at least for the $\Delta=\infty$ Ising limit by droplet states as in \eqref{satexample} above.

\item Part (b) shows that averaging over the disorder eliminates the logarithmic correction and leads to a strict area law. 

\item The proof of Theorem~\ref{thm:XXZarealaw} is given in \cite{BeaudWarzel2017b} for a slightly smaller energy range, based on a Combes-Thomas-type bound similar to the one in Theorem~\ref{thm:CTbound} discussed in Section~\ref{sec:CT} below (and it extends to the energy range stated here with the availability of the CT bound for this larger range). An additional tool used in this context in \cite{BeaudWarzel2017b} is an extension of Theorem~\ref{thm:evdecay} from eigenvectors to eigenprojections, i.e., a bound of the form \eqref{evdecay} for $\|\chi_{\mathcal A} P_{I_\delta}(H)\|$ (Lemma~3.1 in \cite{BeaudWarzel2017b}).

\item The deterministic part (a) follows in a relatively straightforward manner from the eigenprojection bound, using Renyi entropies as a convenient upper bound to the von Neumann entropy $\mathcal{S}(\cdot)$. However, considerable care is needed in carrying out the high-dimensional summations which appear.  

\item Part (b) follows by adapting the strategy of the proof of Part (a), using in addition that in the presence of the positive random field large down spin clusters rarely have energy below $E_2$ (essentially a consequence of large deviations for the sum $\sum_{j=1}^N \omega_j$ of i.i.d.\ random variables, resulting in exponential decay in $N$ of the $N$-particle integrated density of states $\E(\langle \delta_x, P_{I_\delta} (H_N^L(\omega)) \delta_x \rangle)$, $x\in \mathcal{V}_N^L$).

\end{itemize}

\vspace{.3cm}

In order to establish other MBL manifestations in the droplet regime, we will need stronger assumptions on the model than in Theorem~\ref{thm:XXZarealaw}, giving some additional credence to the hierarchy proposed in \eqref{hierarchy} above, according to which an area law for the entanglement should be considered as the weakest form of MBL manifestations considered here.

For the sake of our presentation here, we will strengthen the assumption on the model by requiring that $\Delta$ is sufficiently large (not just $\Delta>1$ as for Theorems~\ref{thm:evdecay} and \ref{thm:XXZarealaw}), putting us in a {\it semi-classical regime} of the associated Schr\"odinger-type operators \eqref{HNL} for the remainder of this paper. Alternatively, the results presented below also hold in a {\it large disorder regime}, i.e., when considering a random field $\lambda \sum_j \omega_j \mathcal{N}_j$ with coupling $\lambda >>0$ (and any $\Delta>1$). We refer to \cite{EKS2017a, EKS2017b} for detailed statements in this regime (in fact, the two regimes can be combined into a joint $(\lambda,\Delta)$-regime, essentially requiring that $\lambda \sqrt{\Delta-1}$ is sufficiently large).

Before proceeding to zero-velocity LR bounds and exponential clustering of correlations, additional tools have to be developed. Central among these is the following result from \cite{EKS2017a}.

\begin{thm}[Droplet Localization] \label{thm:droploc}
Let $\delta>0$ and assume that $\Delta>1$ is sufficiently large. Then there exist $C<\infty$ and $m>0$, depending on $\delta$ and $\Delta$ such that
\begin{equation} \label{1}  
\E\left( \sum_{E\in \sigma(H)\cap I_{\delta}} \|{\mathcal N}_j \psi_E\| \|{\mathcal N}_k \psi_E\| \right) \le C e^{-m|j-k|} 
\end{equation}
uniformly in $L>0$ and $j,k \in [-L,L]$.
\end{thm}

\vspace{.3cm}

Here we use that fact that the spectrum of $H=H_{XXZ}^L(\omega)$ almost surely consists of only non-degenerate eigenvalues (which follows by adapting an argument from Appendix A in \cite{AbdulRahmanStolz2015}). Thus we can denote by $\psi_E$ the unique normalized eigenvector to each $E\in \sigma(H)$.

Remarks:

\begin{itemize}

\item Note that $\|{\mathcal N}_j \psi_E\|^2 = \langle \psi_E, {\mathcal N}_j \psi_E \rangle$ is the expected value of the probability that the eigenstate $\psi_E$ has a down-spin at site $j$. Thus the exponential decay of \eqref{1} in $|j-k|$ can be interpreted as saying that eigenstates to energies in $I_{\delta}$ typically (after disorder averaging) do not have down-spins at widely separated sites. In other words, eigenstates essentially have only a single cluster (droplet) of down-spins. It is in this sense that the one-dimensional degree of freedom $x_1$ of the states in the droplet spectrum of the free XXZ chain (compare the discussion following Theorem~\ref{thm:evdecay}), becomes localized by adding of disorder. It is this what we refer to as ``droplet localization''.

\item On can think of droplet localization \eqref{1} as a form of many-body eigencorrelator localization in the droplet spectrum, compare with the corresponding one-body concept \eqref{ecorloc} above. In particular, note that
\begin{equation} \label{MBeigcor}
\E \left( \sup_{|g|\le \chi_{I_{\delta}}} \|{\mathcal N}_j g(H) \mathcal{N}_k\|_1 \right) \le \E\left( \sum_{E\in \sigma(H)\cap I_{\delta}} \|{\mathcal N}_j \psi_E\| \|{\mathcal N}_k \psi_E\| \right),
\end{equation}
which follows due to almost sure simplicity of eigenvalues, so that $\|{\mathcal N}_j P_E {\mathcal N}_k\|_1 = \|{\mathcal N}_j \psi_E\| \|{\mathcal N}_k \psi_E\|$ for the spectral projection $P_E$ onto $E$ (here $\|\cdot\|_1$ denotes the trace class norm). For the special case $g(H) = P_{I_\delta}(H) e^{-itH}$ this combines with \eqref{1} to
\begin{equation}
\E \left(\sup_{t\in \R} \|{\mathcal N}_j P_{I_\delta}(H) e^{-itH} {\mathcal N}_k\|_1 \right) \le C e^{-m|j-k|},
\end{equation}
which indicates how the results on {\it dynamical} many-body localization to be discussed below follow from \eqref{MBeigcor}.

\item Some ideas behind the proof of Theorem~\ref{thm:droploc} are discussed in Section~\ref{proofdroploc}.
\end{itemize}

We now discuss several consequences of droplet localization, completing the list of prototypical MBL manifestations of the XXZ chain started with Theorem~\ref{thm:XXZarealaw} and similar to the properties of the disordered XY chain discussed in Section~\ref{sec:XY}. All of these will be statements about dynamical MBL in the droplet spectrum, meaning that we have to properly restrict the Heisenberg dynamics $\tau_t(X) = e^{itH} X e^{-itH}$ of an observable $X$ under $H$ to the droplet energy regime. We will do this by considering energy restricted observables  $X_{I_\delta} := P_{I_\delta}(H) X P_{I_\delta}(H)$ and their dynamics
\begin{equation}
\tau_t(X_{I_\delta}) = e^{itH} P_{I_\delta} X P_{I_\delta} e^{-itH}= P_{I_\delta} \tau_t(X) P_{I_\delta} = (\tau_t(X))_{I_\delta}.
\end{equation}
We think of this as the Heisenberg analogue of how one usually restricts the Schr\"odinger dynamics of a state $\psi$ to an energy window $I$ via considering $e^{-itH} P_I(H) \psi$.

We recall that an observable $X$ on $\mathcal{H}_{[-L,L]} = \bigotimes_{j=-L}^{L} \C^2$ is said to have support $S_X=[a,b]\subset [-L,L]$ if it is of the form $X = \idty_{left} \otimes A \otimes \idty_{right}$, where $A$ is an observable on $\mathcal{H}_{[a,b]}$ and $\idty_{left}$ and $\idty_{right}$ are the identity operators on $\mathcal{H}_{[-L,a-1]}$ and $\mathcal{H}_{[b+1,L]}$, respectively.

We start with exponential decay of correlations:

\begin{thm}[Exponential clustering] \label{thm:XXZcluster}
If $\delta>0$ and $\Delta>1$ is sufficiently large (depending on $\delta$), then there exist $C<\infty$ and $m>0$ such that for all local observables $X$ and $Y$,
\begin{equation} \label{eq:XXZcluster}
\E \left( \sup_{t\in \R} \sum_{E\in \sigma(H)\cap {I_\delta}} \mbox{Cor}_{\psi_E}(\tau_t(X_{I_\delta}), Y_{I_\delta}) \right) \le C\|X\| \|Y\| e^{-m \,{\rm dist}(S_X, S_Y)}.
\end{equation}
\end{thm}

For convenience (and consistence of presentation within this article) we state this result in a form given in \cite{EKS2017b}. A closely related and slightly stronger variant of this result was already proven in \cite{EKS2017a} (using a slightly different way to carry out the energy restriction to $I_\delta$ in the definition of the correlation).

As in Theorem~\ref{thmexpclust} for the disordered XY chain, time-dependent correlations are considered here. 
We point out that this form of exponential clustering is not only uniform in time $t$ and in all eigenstates to energies in the droplet spectrum, but holds for the {\it sum} of the correlations of all these states. This is, of course, a consequence of the summation over all such states allowed in the droplet localization bound \eqref{1} (or, viewed alternatively, the fact that the left hand side of \eqref{MBeigcor} allows for the use of the trace class norm, not just the operator norm).

In Section~\ref{proofexpcl} below we will provide a proof of a special case of this result, considering only time $t=0$ and observables $X$ and $Y$ supported at a single site. This is not very difficult and provides a prototypical argument for how some of the relevant MBL manifestations can be derived from droplet localization \eqref{1}. The following two theorems on zero-velocity LR bounds and the related quasi-locality of the Heisenberg dynamics are the two main results of \cite{EKS2017b} and require more work (which one could see as another confirmation of the hierarchy \eqref{hierarchy}, at least as a difficulty ranking). They also derive from \eqref{1}, but we will not get into any details of their proofs here. 

\begin{thm}[Zero-velocity LR bound] \label{thm:XXZLR}
Let $\delta>0$ and $\Delta>1$ sufficiently large. Then there exist $C<\infty$ and $m>0$ such that, for all $L\in \N$ and observables $X$ and $Y$ on $\mathcal{H}_{[-L,L]}$ supported on $S_X$ and $S_Y$, respectively,
\begin{equation} \label{eq:XXZLR}
\E \left( \sup_{t\in \R} \| [ \tau_t(X_{I_\delta}), Y_{I_\delta} ] \|_1 \right) \le C \|X\| \|Y\| e^{-m \,{\rm dist}(S_X, S_Y)}.
\end{equation}
\end{thm}

This LR bound requires energy restriction to the interval $I_\delta$ in the droplet spectrum. One might naturally think that given the trivial product structure of the all-spins-up ground state $\Omega$, Theorem~\ref{thm:XXZLR} remains true if one instead restricts the energy to $I_\delta \cup \{0\}$ (which, given the gap between ground state energy and droplet spectrum, is the same as restricting to $I_{0,\delta} := \left[0, (2-\delta) \big(1-\frac{1}{\Delta} \big)\right]$. However, as was found in \cite{EKS2017b}, this is not true! When attempting to prove this one has to include non-trivial ``counter terms'' on the left hand side of \eqref{eq:XXZLR}, reflecting an interaction between the groud state and the droplet states in the dynamics. 

On the other hand, it is possible to prove the following {\it double commutator} LR bound for the larger interval $I_{0,\delta}$, involving three observables $X$, $Y$ and $Z$:
\begin{eqnarray}
\lefteqn{\E \left( \sup_{t,s \in \R} \| [[ \tau_t(X_{I_{0,\delta}}), \tau_s(Y_{I_{0,\delta}}) ], Z_{I_{0,\delta}}] \|_1 \right) } \\
& \le & C \|X\| \|Y\| \|Z\| e^{-m \min\{ \rm{dist}(S_X,S_Y), \rm{dist}(S_X,S_Z), \rm{dist}(S_Y, S_Z)\}}, \notag
\end{eqnarray}
see Theorem~2.3 in \cite{EKS2017b}.

Finally, we mention a result on quasi-locality of the Heisenberg dynamics for energies in the droplet spectrum. It says that the support of an observable $X$ is essentially unchanged under the dynamics, up to an exponentially small quantum tail. As in the earlier results this requires both-sided projection onto the droplet spectrum.

\begin{thm}[Quasi-locality of the dynamics] \label{thm:XXZdyn}
Let $\delta>0$ and $\Delta>1$ sufficiently large. Then there exist $C<\infty$ and $m>0$ with the following property: For all $L\in \N$, observables $X$ in $\mathcal{H}_{[-L,L]}$ with support $S_X$, $t\in \R$ and $\ell \in \N$, there exists an observable $X_{\ell}(t) = X_{\ell}(t,\omega)$ supported on 
\begin{equation} 
S_{X,\ell} := (S_X +[-\ell,\ell]) \cap [-L,L]
\end{equation}
and such that
\begin{equation} 
\E \left( \sup_{t\in \R} \| (X_{\ell}(t) - \tau_t(X))_{I_\delta} \| \right) \le C \|X\| e^{-m\ell}.
\end{equation}
\end{thm}

In situations where no energy restriction is needed, Lieb-Robinson bounds and quasi-locality of the dynamics are equivalent, see \cite{NOS2006} where this is discussed in the more general setting of spin systems with finite positive LR velocity. The non-trivial part of this equivalence (from LR bounds to quasi-locality) requires integration with respect of Haar measure on a group of unitary operators. This does not carry over to the energy-restricted versions of these properties discussed here (as previously observed in \cite{Friesdorfetal2015}), so that Theorems~\ref{thm:XXZLR} and \ref{thm:XXZdyn} require separate proofs, although related by both starting from Theorem~\ref{thm:droploc}, see \cite{EKS2017b} for details.

\section{Some illustrative proofs} \label{sec:proofs}

The plan of this section is to 

\begin{itemize}

\item State the Combes-Thomas bound behind Theorem~\ref{thm:evdecay} on the droplet structure of eigenstates, sketch its proof, and show how 
Theorem~\ref{thm:evdecay} follows from it.

\item Prove the  special case of $t=0$ and one-site observables of Theorem~\ref{thm:XXZcluster} on exponential clustering.

\item Sketch some of the ideas behind the proof of droplet localization, Theorem~\ref{thm:droploc}.

\end{itemize}
We admit that for these illustrations of the methods we have avoided going into some of the technically most challenging parts of the arguments.

\subsection{A Combes-Thomas bound and the proof of Theorem~\ref{thm:evdecay}} \label{sec:CT}

Here we will outline a proof of Theorem~\ref{thm:evdecay}. For ease of presentation we will do this in infinite volume, thus avoiding to drag around boundary conditions, i.e., for the operator
\begin{eqnarray} \label{Hinf}
H \:=\: H_N & = & -\frac{1}{2\Delta} A + \frac{1}{2}D + V \\
& = & -\frac{1}{2\Delta} {\mathcal L} + \frac{1}{2}\Big(1-\frac{1}{\Delta}\Big) D + V \notag
\end{eqnarray}
on $\ell^2({\mathcal V}_N)$. Here $N\in \N$ is kept fixed throughout and will mostly be dropped from the notation. Here $A$ is the adjacency operator, $L = A - D$ graph Laplacian and $V\ge 0$ any non-negative potential on $\ell^2({\mathcal V}_N)$. (The arguments below work in the same way for the finite volume operators \eqref{HNL}, with constants not depending on the volume and boundary condition, as long as \eqref{beta} is assumed, giving the required positivity properties.)

Recall that $D(x)$ is the degree of $x$ in the graph $\mathcal{V}_N$ (or twice the number of connected components in the ordered configuration $x=(x_1,\ldots,x_N)$. 

By $P_1$ we denote the orthogonal projection onto $\ell_2({\mathcal V}_{N,1})$ (the subspace spanned by the droplet configurations) in $\ell_2({\mathcal V}_{N})$. Using  \eqref{Hinf}, we get from $L\ge 0$, $V\ge 0$, $D(x)=2$ for $x\in \mathcal{V}_{N,1}$ and $D(x)\ge 2$ for $x\in \mathcal{V}_N \setminus \mathcal{V}_{N,1}$ that 
\begin{equation} \label{Hlowbound}
H+\Big(1-\frac{1}{\Delta}\Big)P_1 \ge 2\Big(1-\frac{1}{\Delta}\Big),
\end{equation}
so that by known methods the Schr\"odinger-type operator $H+(1-1/\Delta)P_1$ will satisfy a Combes-Thomas bound for energies in the droplet spectrum. The important fact, however, is that we can show this with constants independent of $N$ and with exponential decay in the $N$-dimensional $\ell^1$-distance:

\begin{thm}[Combes-Thomas bound] \label{thm:CTbound}
Let $\delta>0$, $E \le (2-\delta)(1-\frac{1}{\Delta})$ and ${\mathcal A}, {\mathcal B} \subset {\mathcal V}_N$. Then
\begin{equation} 
\| \chi_{\mathcal A} \left(H+\Big(1-\frac{1}{\Delta}\Big)P_1-E\right)^{-1} \chi_{\mathcal B}\| \le \frac{16\Delta}{\delta(\Delta-1)} \left( 1 + \frac{\delta(\Delta-1)}{8} \right) ^{-d_N({\mathcal A}, {\mathcal B})} .
\end{equation}
\end{thm}

\vspace{.3cm}

Variants of Theorem~\ref{thm:CTbound} have been provided in \cite{EKS2017a} and \cite{BeaudWarzel2017a}. The presentation below is close to how the argument was reproduced in \cite{FischbacherStolz2017} (for XXZ models on more general graphs).

We mention that for the applications in Section~\ref{MBLXXZ} one needs some extensions and variations of Theorem~\ref{thm:CTbound}. First, one can work with complex energy $E+i\epsilon$. Also, instead of lifting the spectral minimum of $H$ above $E_2$ by adding $(1-1/\Delta)P_1$, one can prove a very similar Combes-Thomas bound by restricting the Hamiltonian $H$ to the range $\ell^2(\mathcal{V}_N \setminus \mathcal{V}_{N,1})$ of $\bar{P} = \idty-P_1$, i.e., the non-droplet configurations (a situation considered in both \cite{EKS2017a} and \cite{BeaudWarzel2017a}).

\vspace{.3cm}
\begin{proof} As in the ``standard proof'' of Combes-Thomas bounds (e.g.\ \cite{Kirsch}) we use dilations:
\begin{equation}
K_{\eta}:= e^{-\eta \rho_{\mathcal A}} H e^{\eta \rho_{\mathcal A}} -H,
\end{equation}
where $\eta>0$ and $\rho_{\mathcal A}(x) := d_N({\mathcal A},x)$. A calculation shows
\begin{equation} \label{Keta}
(K_{\eta} \psi)(x) = \frac{1}{2\Delta} \sum_{y\in {\mathcal V}_N: y \sim x} (1- e^{-\eta (\rho_{\mathcal A}(y)-\rho_{\mathcal A}(x)}) \psi(y),
\end{equation}
where $x\sim y$ denotes next neighbors. For these we have $|\rho_{\mathcal A}(y)-\rho_{\mathcal A}(x)| \le 1$ and thus
\begin{equation}
|1- e^{-\eta (\rho_{\mathcal A}(y)-\rho_{\mathcal A}(x)}| \le e^{\eta} -1
\end{equation}
In the standard proof one would now conclude that $\|K_{\eta}\| \le \frac{e^{\eta}-1}{2\Delta} \cdot 2N$ (as $2N$ is the maximal degree of ${\mathcal V}_N$). But this is not good enough for our purposes as it would lead to an exponential decay rate proportional to $1/N$ in the Combes-Thomas bound.

Instead one ``borrows'' two factors $D^{-1/2}$, starting with $(D^{-1/2} K_\eta D^{-1/2} \psi)(x)$ in \eqref{Keta} and gets from a slightly more careful calculation (compare \cite{EKS2017a} or \cite{FischbacherStolz2017}) that
\begin{equation} \label{Dminus}
\|D^{-1/2} K_{\eta} D^{-1/2}\| \le \frac{e^{\eta}-1}{\Delta}.
\end{equation}
That this bound is independent of $N$ is a consequence of the ``balance'' of the operators $A$ and $D$ in \eqref{Hinf}, i.e., that at each site $x$ the number of hoppings emanating from $x$ is the degree $D(x)$ of the graph (in fact, $D$ dominates $A$ due to $\Delta>1$).

The two borrowed factors can be ``paid back'', using ${\mathcal L}\ge 0$ and quadratic forms to get (with $R_E := (H+(1-1/\Delta) P_1 -E)^{-1}$)
\begin{equation} \label{Dplus}
\left\|D^{1/2}  R_E D^{1/2}\right\| \le  \left\|D^{1/2} \Big(\Big( 1-\frac{1}{\Delta} \Big) \Big(\frac{1}{2}D+P_1 -(2-\delta)\Big)\Big)^{-1} D^{1/2}\right\|  \le  \frac{4\Delta}{\delta(\Delta-1)}. 
\end{equation}
(Consider $x\in \mathcal{V}_{N,1}$ and $x\in \mathcal{V}_N \setminus \mathcal{V}_{N,1}$ separately for the multiplication operator in the last step.) Again, the most relevant fact about the bound \eqref{Dplus} is the independence of $N$.

At this point one can return to the (essentially) standard proof, bounding
\begin{eqnarray}
\| \chi_{\mathcal A} R_E \chi_{\mathcal B}\| & \le & \| \chi_{\mathcal A} D^{1/2} R_E D^{1/2} \chi_{\mathcal B}\| \\
& \le & \| \chi_{\mathcal A} e^{\eta \rho_{\mathcal A}}\| \| D^{1/2} e^{-\eta \rho_{\mathcal A}} R_E e^{\eta \rho_{\mathcal A}} D^{1/2}\| \|e^{-\eta \rho_{\mathcal A}} \chi_{\mathcal B} \|. \notag
\end{eqnarray}
The first term is $1$, the third term provides the exponential decay, and the second term can be shown to be bounded as a consequence of \eqref{Dminus} and \eqref{Dplus} when choosing 
\begin{equation} \label{gamma}
\eta = \log(1+\delta(\Delta-1)/8).
\end{equation} 
For more details again see  \cite{EKS2017a} or \cite{FischbacherStolz2017}.
\end{proof}

It is now straightforward to conclude Theorem~\ref{thm:evdecay} from this (using the finite volume version of the Combes-Thomas bound): For a normalized eigenvector $\psi$ of $H_N$ to energy $E\in I_\delta$ and $\mathcal{A} \subset \mathcal{V}_N^L$ one has, with $\gamma$ from \eqref{gamma},
\begin{eqnarray}
\| \chi_{\mathcal A} \psi \| & = &  \left\| \chi_{\mathcal A} R_E (H_N+\big(1-\frac{1}{\Delta}\big)P_1-E) \psi \right\| \\
& = & \big(1-\frac{1}{\Delta}\big) \| \chi_{\mathcal A} R_E \chi_{\mathcal{V}_{N,1}} \psi \| \notag \\
& \le & \frac{16}{\delta}  e^{-\eta d_N(\mathcal{A}, \mathcal{V}_{N,1})} \| \chi_{\mathcal{V}_{N,1}} \psi\|. \notag
\end{eqnarray}

\subsection{Exponential Clustering} \label{proofexpcl}

Our goal here is to derive a special case of Theorem~\ref{thm:XXZcluster} from Theorem~\ref{thm:droploc}. We will only consider the case of $t=0$ and of  one-site observables $X\in {\mathcal A}_j$, $Y\in {\mathcal A}_k$, $j\not= k$. 

By linearity, it suffices to consider 16 cases, i.e., where $X$ is given by one of the four observables
\begin{equation} 
X^{+,+} = \begin{pmatrix} 1 & 0 \\ 0 & 0 \end{pmatrix}_{\hspace{-0.1cm} j}, \: X^{+,-} = \begin{pmatrix} 0 & 1 \\ 0 & 0 \end{pmatrix}_{\hspace{-0.1cm} j}, \: X^{-,+} = \begin{pmatrix} 0 & 0 \\ 1 & 0 \end{pmatrix}_{\hspace{-0.1cm} j}, \: X^{-,-} = \begin{pmatrix} 0 & 0 \\ 0 & 1 \end{pmatrix}_{\hspace{-0.1cm} j}
\end{equation}
and $Y$ by one of the four observables $Y^{\pm,\pm}$, defined similarly at site $k$.

We start with the case $X^{-,-} = {\mathcal N}_j$ and $Y^{-,-} = {\mathcal N}_k$, in which, for every $E\in I_\delta$,
\begin{equation}
\mbox{Cor}_{\psi_E}(X_{I_\delta}, Y_{I_\delta}) = |\langle \psi_E, \mathcal{N}_j P_{I_\delta} (\idty - |\psi_E \rangle \langle \psi_E|) P_{I_\delta} \mathcal{N}_k \psi_E \rangle| \le \| \mathcal{N}_j \psi_E \| \|\mathcal{N}_k \psi_E\|.
\end{equation}
Thus in this case \eqref{eq:XXZcluster} follows directly from Theorem~\ref{thm:droploc}.

Several other cases are trivial due to particle number conservation: Assume that $\psi = \psi_E \in {\mathcal H}_N$ for some $N$ and all eigenvalues $E$, which holds almost surely by simplicity of the spectrum of $H_N$. Then, if either $Z=X$ or $Z=Y$ it holds that
\begin{equation} 
Z^{+,-} \psi \:\mbox{has at most $N-1$ particles, i.e., lies in $\bigoplus_{j=0}^{N-1} {\mathcal H}_j$},
\end{equation}
\begin{equation}
Z^{-,+} \psi \:\mbox{has at least $N+1$ particles, i.e., lies in $\bigoplus_{j=N+1}^{2L+1} {\mathcal H}_j$}.
\end{equation}
Furthermore, $Z^{-,-}$ and $P_I$ preserve the particle number, i.e., they leave all the spaces $\mathcal{H}_j$ invariant.
This settles five more cases, for example
\begin{eqnarray}
\mbox{Cor}_\psi(X^{+,-}_{I_\delta}, Y^{+,-}_{I_\delta} ) & = & | \langle \psi, X^{+,-}, P_{I_\delta} Y^{+,-} \psi \rangle - \langle \psi, X^{+,-} \psi \rangle \langle \psi, Y^{+,-} \psi \rangle| \\
& = & |\langle X^{-,+} \psi, P_{I_\delta} Y^{+,-} \psi \rangle| = 0,
\end{eqnarray} 
and similarly for the cases $(X^{-,+},Y^{-,+})$, $(X^{+,-}, Y^{-,-})$, $(X^{-,-}, Y^{-,+})$ and $(X^{-,-}, Y^{+,-})$. 

Eight of the remaining ten cases can be reduced to the previous six cases by using the properties
\begin{equation} 
\mbox{Cor}_\psi(X^{+,+}_{I_\delta}, Z_{I_\delta}) = \mbox{Cor}_\psi(X^{-,-}_{I_\delta}, Z_{I_\delta} ) \quad \mbox{(use that $X^{+,+} = I-X^{-,-}$)},
\end{equation}
\begin{equation} 
\mbox{Cor}_\psi( Z_{I_\delta}, Y^{+,+}_{I_\delta} ) = \mbox{Cor}_\psi(Z_{I_\delta}, Y^{-,-}_{I_\delta}),
\end{equation}
\begin{equation} 
\mbox{Cor}_\psi(Z,W) = \mbox{Cor}_\psi(W^*, Z^*).
\end{equation}

After doing the necessary bookkeeping, one is left with only two cases to be considered, the combinations $(X^{-,+}, Y^{+,-})$ and $(Y^{-,+}, X^{+,-})$. The second of these reduces to the first by commutation. The remaining case is settled by
\begin{eqnarray}
\sum_E \mbox{Cor}_{\psi_E}(X^{-,+}_{I_\delta},Y^{+,-}_{I_\delta}) & = & \sum_E |\langle \psi_E, X^{-,+} P_{I_\delta} (\idty-P_E) Y^{+,-} \psi_E \rangle| \\
& = & \sum_E |\langle \psi_E, {\mathcal N}_j X^{-,+} P_{I_\delta} (\idty-P_E) Y^{+,-} {\mathcal N}_k \psi_E \rangle| \notag \\
& \le & \sum_E \|{\mathcal N}_j \psi_E\| \|{\mathcal N}_k \psi_E\| \notag,
\end{eqnarray}
so that we can once again apply Theorem~\ref{thm:droploc} to conclude.

\subsection{On the proof of Theorem~\ref{thm:droploc}} \label{proofdroploc}

As far as localization properties (and their proofs) are concerned, the proof of Theorem~\ref{thm:droploc} represents the technical core of all the MBL manifestations derived from it, i.e., Theorems~\ref{thm:XXZcluster}, \ref{thm:XXZLR} and \ref{thm:XXZdyn}. Its proof comprises a thorough adaptation of techniques from the theory of random Schr\"odinger, applied to the $N$-particle restrictions $H_N$, and developing the means necessary to achieve bounds which are uniform in the particle number. At the center of the argument in \cite{EKS2017a} (and also some closely related results in \cite{BeaudWarzel2017a}) are tools based on the fractional moments method. Here, for the sake of brevity and to keep this account from getting highly technical, we mostly limit ourselves to explaining how one gets started. A somewhat more thorough outline of the ideas and challenges behind the proof is provided in Section 2 of \cite{EKS2017a}.

\begin{itemize}
\item Particle number conservation allows to write \eqref{1} in the form of a statement of localization for suitable eigenstate correlators (at least formally resembling \eqref{ecorloc}) of the operators $H_N$. For this let $Q_{j,N}$ denote the restriction of the (particle number conserving) $\mathcal{N}_j$ to the $N$-particle subspace $\mathcal{H}_N = \ell^2(\mathcal{V}_N^L)$. This turns out to be the characteristic function of the set
\begin{equation}
S_{j,N} = \{x\in \mathcal{V}_N^L: x_k = j \:\mbox{for some} \: k \in [1,N]\},
\end{equation}
i.e., the set of configurations at which the random potential depends on $\omega_j$.
If we now define $N$-body eigenstate correlators by 
\begin{equation} 
Q_N(j,k;I_\delta) = \sum_{E\in \sigma(H_N) \cup I_\delta} \|Q_{j,N} P_E Q_{j,N}\|_1,
\end{equation}
where $P_E$ denotes the spectral projection of $H_N$ on $E$, then we can decompose the droplet correlations into their $N$-particle contributions,
\begin{equation}
\sum_{E\in \sigma(H)\cap I_{\delta}} \|{\mathcal N}_j \psi_E\| \|{\mathcal N}_k \psi_E\| = \sum_{N=1}^\infty Q_N(j,k;I_\delta),
\end{equation}
so that we are left with finding exponentially decaying bounds for $\E( Q_N(j,k;I_\delta))$, which must also be summable in $N$.

\item Bounds on correlators of the form $\E (Q_N(j,k;I_\delta))$ are, at least in principle, within the range of results which can be obtained through a Green's function analysis via the fractional moments method (\cite{AizenmanMolchanov, AizenmanWarzel, Stolz}). The crucial difficulty which arises is that these correlators are long range correlated in $j$ and $k$, in that the sets $S_{j,N}$ and $S_{k,N}$ overlap within the ``bulk region'' $\mathcal{V}_N^L \setminus \mathcal{V}_{N,1}^L$, thus also causing correlations of unbounded length in the random parameters $\omega_j$ and $\omega_k$.

However, as explained by \eqref{Hlowbound}, energies in $I_\delta$ are classically accessible only through the droplet configuations $\mathcal{V}_{N,1}^L$ and, when restricted to this one-dimensional subspace of the configuration space, the sets $S_{j,N}$ for adjacent indices $j$ overlap only in $N$ neighboring points. 

At this point a two-part strategy for showing decay bounds on the Green functions of the operators $H_N$ emerges: (i) Control the Green function in the bulk $\mathcal{V}_N^L \setminus \mathcal{V}_{N,1}^L$ by showing exponential decay in this classically forbidden region via a Combes-Thomas bound similar to \eqref{thm:CTbound} (for independence of the bounds on $N$ it is crucial that the exponential decay holds in the $\ell^1$-distance of $N$-particle configurations), (ii) Carry out an essentially one-dimensional fractional moment analysis of the Green function along the ``edge'' $\mathcal{V}_{N,1}^L$ of the configuration space. It is here more than anywhere else in the proof that it becomes apparent how proving droplet localization in the XXZ chain is reduced to proving one-dimensional localization for the quasi-particle given by a spin droplet.

\item After all is said and done this leads to an exponential decay bound for $\E (Q_N(j,k;I_\delta))$, which is uniform in $N$. That one also gets summability in $N$ is, just as discussed following Theorem~\ref{thm:XXZarealaw}, ultimately a consequence of a large deviations argument in $N$, i.e., the fact that for large particle number the $N$-body random potential will shift the energy of $H_N$ above the droplet spectrum with high probability.

\end{itemize}

\section{Epilogue: Where to go from here?} \label{sec:problems}

There is no denying that the mathematical theory of many-body localization is still in its infancy, including for the relatively accessible case of quantum spin systems. Up to this point the available fully rigorous results have focused on specific models, where special features are available which allow for some highly tailored arguments. We hope that within another decade the understanding of MBL will have grown sufficiently to provide a more comprehensive understanding of classes of disordered quantum spin systems, in particular for the one-dimensional case. Nevertheless, the study of specific examples, with the aim of gradually approaching a more universal picture, will continue to be important.

We conclude this paper with a (not very systematic) list of open questions and directions which might be worthwhile pursuing.

\begin{itemize}

\item The most interesting remaining open problem related to the disordered XY chain is probably the understanding of entanglement properties of its {\it thermal states}. These are mixed states, so the bipartite entanglement entropy \eqref{entangle} is not an appropriate entanglement measure. Among the more suitable concepts is the logarithmic negativity wit respect to a bipartite decomposition, e.g.\ \cite{VidalWerner2002} and \cite{Plenio2005}. However, this involves taking partial transposes (rather than partial traces), an operation which does {\it not} preserve the property of a state to be quasi-free. Finding a way to overcome this issue is the main challenge here.

\item Many questions remain open for the disordered XXZ model, even in the Ising phase (not to speak of the gapless Haldane phase which appears for $\Delta \le 1$). The physically desirable ultimate goal would be to show that MBL properties hold in an {\it extensive} energy regime at the bottom of the spectrum, i.e., in an interval which grows as $\rho L$ in the length $L$ of the chain, with a positive energy density $\rho>0$. So far, the results presented in Section~\ref{MBLXXZ} only hold for the fixed size droplet band $I=[1-1/\Delta, 2(1-1\Delta)]$, which can be physically interpreted as a form of {\it zero-temperature localization}. One might also think of this as a many-body analogue of the Lifshitz tail regime in the Anderson model (with ``thinness'' of the spectrum in a fluctuation boundary regime substituted by smallness of the $N$-particle density of states). 

One interesting question to pursue is the fate of MBL  in the XXZ chain in the $k$-cluster bands $[k(1-1/\Delta), (k+1)(1-1/\Delta)]$, $k=2,3,\ldots$, where the down-spin droplets of the free XXZ chain break up into $k$ connected components of down-spins (a fact already confirmed by an extension of the Combes-Thomas estimates provided in \cite{EKS2017a} and \cite{FischbacherStolz2017} to these higher energy bands). Studying the effect of disorder on the $k$-cluster bands can be understood as understanding how $k$ down-spin droplets (given by the $k$ clusters) interact in a disordered exterior field. We find it likely that some MBL properties will still hold, in particular for a sufficiently strong semi-classical regime $\Delta>>1$ or for fields $\sum_j \omega_j \mathcal{N}_j$ with random variables of sufficiently large disorder. However, the form of droplet localization described in Theorem~\ref{thm:droploc} will {\it not} hold in these higher bands, as has been verified in Theorem~2.1 of \cite{EKS2017b}. New ideas, potentially with a more scattering theoretic flavor, are needed. In particular, when interpreting the $k$ clusters as $k$ interacting quasi-particles in random environment, then existing results and methods for the multi-particle Anderson model (e.g.\ \cite{ChulaevskySuhov1, ChulaevskySuhov2,AizenmanWarzelNbody}) may provide some useful guidance.

Note that proving MBL properties in the $k$-cluster bands, for any given $k$, extends zero-temperature localization to higher energies, but does still not give an extensive energy regime $\rho L$ in the limit of large systems. Approaching a proof of  the latter seems to need an entirely different strategy.

\item Another interesting question for the XXZ chain is if the above results extend to the case where a slight anisotropy is introduced into the XX term of the interaction, similar to what we discussed for the XY chain in Section~\ref{sec:anisoXY}. If the anisotropy parameter is sufficiently small (and $\Delta>1$), then methods such as those presented in \cite{MoonNachtergaele} show that the ground state gap remains open (uniformly in the volume). In fact, for the free spin Hamiltonian and $\Delta$ sufficiently large there will still be another gap separating the ``droplet band'' from the higher energy bands above it. That one can still speak of a droplet band in this anisotropic setting would need more justification, in showing suitable properties of the eigenstates to energies in this band.

The most difficult (and most interesting) challenge here would be to do this despite the fact that the anisotropic model does not conserve the particle number any longer. Thus the convenient decomposition of the XXZ Hamiltonian into the $N$-particle operators $H_N$, quite crucial for everything we did above, does not apply and the concept of droplets (as well as their localization after adding disorder) would have to be suitably described and proved in the setting of the full Fock space.

\item There are two other models of many-body systems with disorder, in which localization properties can be rather directly analyzed in terms of an effective one-particle Hamiltonian: Disordered harmonic oscillator systems (e.g.\ \cite{AbdulRahman2017} and references therein) and the Tonks-Girardeau gas \cite{SeiringerWarzel}. A useful feature of harmonic oscillator systems is that they can be studied on multi-dimensional lattices, while the Tonks-Girardeau gas is a model with a continuum configuration space (in fact a continuum analogue of the XY chain). For these reasons, in particular, both models justify further attention.

Another model which should be accessible to the study of MBL properties is the {\it higher spin} XXZ chain, see \cite{Mulherkaretal} for a study in the deterministic setting. This work starts to expose a low energy regime with properties similar to the droplet phase of the spin-$1/2$ chain. The higher spin model is still particle number conserving, but now with the possibility of multiple occupation per site. We find it likely that in the disordered setting at least some of the methods behind the results described in Section~\ref{MBLXXZ} extend to the higher spin case.

Among concrete models we finally mention disordered systems of quantum rotors. An early result on exponential decay of a special ground state correlation for this model has been shown in \cite{KleinPerez1992}, using a multiscale analysis approach. This model has features (including a stable ground state gap in the constant coefficient deterministic case) which should make a revisit interesting, given the new methods in many-body theory which have become available.

\item The best chance for a mathematical theory of MBL to move beyond studying specific models is likely to be the fully many-body localized regime of one-dimensional spin chains. This is physically expected to hold for models with short range interaction and large local disorder, say (to stay with next-neighbor interactions) for a Hamiltonian of the type \eqref{Ham},
\begin{equation}
H = \sum_j h_{j,j+1} + \lambda \sum_j \omega_j t_j,
\end{equation} 
where the disordered local field is coupled by a large parameter $\lambda>>0$. Alternatively, one can look at the case of weak interaction.
Proving suitable forms of MBL at all energies for models of this type would complete the program proposed in \cite{Imbrie2016}, where a model of this type is considered under a physically reasonable assumption on the spacing of the eigenvalues.  Proving such level spacing properties, which can be considered as many-body versions of Wegner or Minami-type estimates, is mathematically far from trivial. We believe that additional tools from many-body theory, such as the control of many-body transport which can be obtained through Lieb-Robinson bounds, will be needed to make further progress on this question.

\item At the very conclusion of this article let us recall some of the topics in the area of MBL which we haven't even touched here and which may currently be out of reach for mathematical investigations. Nevertheless, these questions need to be kept in mind:

\begin{itemize}

\item Can MBL properties be shown for any interesting models of higher-dimensional spin systems? This is not even understood for the  XY model (as one can not map any longer to a free Fermion system). Progress on this will require much more groundwork for deterministic multi-dimensional spin systems, where much less is known than for spin chains.

\item What can be said about MBL in systems of particles other than spins, in particular, interacting particles which have spatial degrees of freedom? This is the topic of the celebrated work \cite{BAA2006}, where the persistence of Anderson localization in weakly interacting disordered electron gases was proposed.

\item What is many-body {\it delocalization} (in physics generally referred to as thermalization)? And, assuming that a delocalized phase exists, is there a many-body analogue of a mobility edge between localized and delocalized regimes? To correctly gauge the difficulty of this task, one should keep in mind that this question is mathematically not even settled for the three-dimensional non-interacting Anderson model.

\end{itemize}

\end{itemize} 

\bibliographystyle{amsplain}

\end{document}